\documentclass[journal]{IEEEtran}

\usepackage{times}
\usepackage{epsfig}
\usepackage{graphicx}
\usepackage{amsmath}
\usepackage{amssymb}
\usepackage{algorithmic}
\usepackage{multirow}
\usepackage[table,xcdraw]{xcolor}

\usepackage[breaklinks=true,bookmarks=false]{hyperref}

\usepackage{caption}
\newcommand{\x}{\times}
\graphicspath{{Figs/}}

\DeclareMathOperator*{\argmax}{argmax}

\begin{document}
	
	\title{Confidence Measure Guided Single Image De-raining}

	\author{Rajeev Yasarla,~\IEEEmembership{Student Member,~IEEE} and 
		Vishal~M.~Patel,~\IEEEmembership{Senior Member,~IEEE}
		\thanks{Rajeev Yasarla is with the Whiting School of Engineering, Johns Hopkins University, 3400 North Charles Street, Baltimore, MD 21218-2608, e-mail: ryasarl1@jhu.edu}
		\thanks{Vishal M. Patel is with the Whiting School of Engineering, Johns Hopkins University, e-mail: vpatel36@jhu.edu}
		\thanks{Manuscript received...}}

	\maketitle

	\begin{abstract}
		Single image de-raining is an extremely challenging problem since the rainy images contain rain streaks which often vary in size, direction and density. This varying characteristic of rain streaks affect different parts of the image differently.
		Previous approaches have attempted to address this problem by leveraging some prior information to remove rain streaks from a single image.   One of the major limitations of these approaches is that they do not consider the location information of rain drops in the image. The proposed Image Quality-based single image Deraining using Confidence measure (QuDeC), network addresses this issue by learning the quality or distortion level of each patch in the rainy image, and further processes this information to learn the rain content at different scales.  In addition, we introduce a technique which guides the network to learn the network weights based on the confidence measure about the estimate of both quality at each location and residual rain streak information (residual map). Extensive experiments on synthetic and real datasets  demonstrate that the proposed method achieves significant improvements over the recent state-of-the-art methods.

	\end{abstract}
	

	\IEEEpeerreviewmaketitle

	\section{Introduction}
	Many practical computer vision-based systems such as surveillance and autonomous driving often require processing and analysis of videos and images captured under adverse weather conditions such as rain, snow, haze etc.  These weather-based conditions adversely affect the visual quality of images and as a result often degrade the performance of vision systems. Hence, it is important to develop algorithms that can automatically remove these artifacts before they are fed to a vision-based system for further processing. 
	
	\begin{figure}[t!]
		\centering
		\includegraphics[width=1.29cm,height=0.98cm]{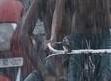}
		\includegraphics[width=1.29cm,height=0.98cm]{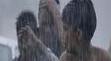}
		\includegraphics[width=1.29cm,height=0.98cm]{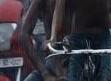}
		\includegraphics[width=1.29cm,height=0.98cm]{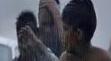}
		\includegraphics[width=1.29cm,height=0.98cm]{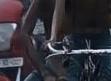}
		\includegraphics[width=1.29cm,height=0.98cm]{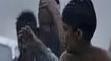}\\
		\includegraphics[width=2.7cm,height=1.58cm]{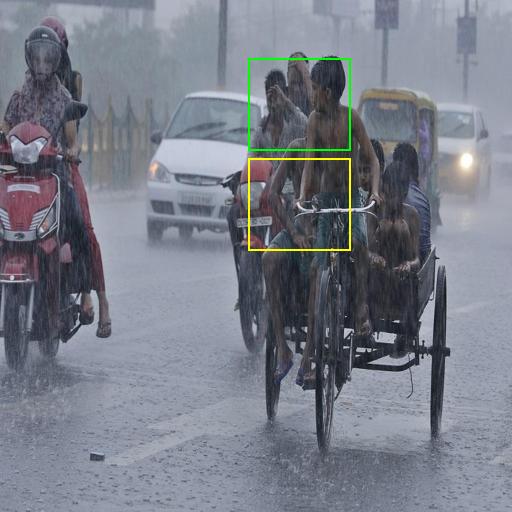}
		\includegraphics[width=2.7cm,height=1.58cm]{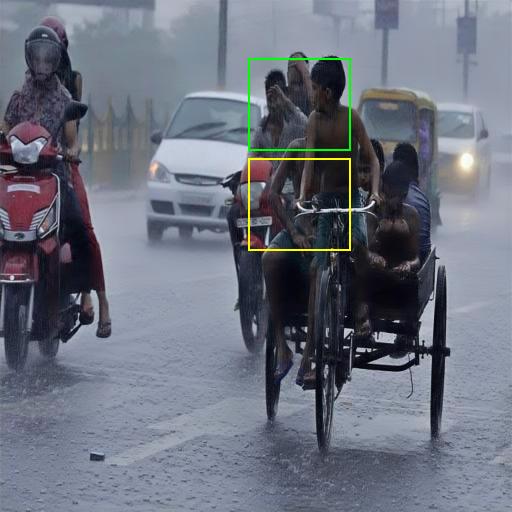}
		\includegraphics[width=2.7cm,height=1.58cm]{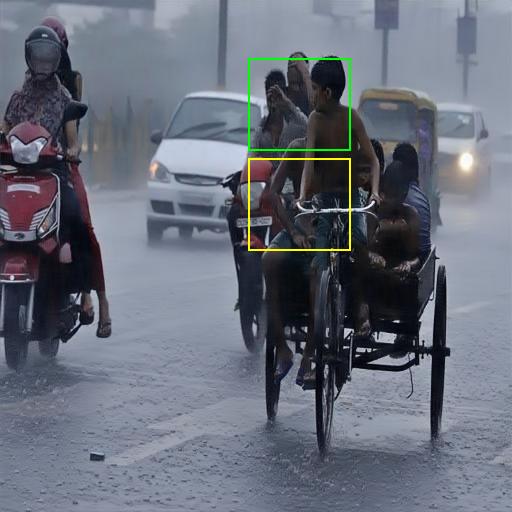}\\
		(a)\hskip70pt(b)\hskip70pt(c)\\
		\includegraphics[width=1.29cm,height=0.98cm]{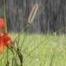}
		\includegraphics[width=1.29cm,height=0.98cm]{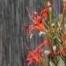}
		\includegraphics[width=1.29cm,height=0.98cm]{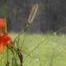}
		\includegraphics[width=1.29cm,height=0.98cm]{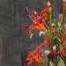}
		\includegraphics[width=1.29cm,height=0.98cm]{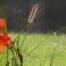}
		\includegraphics[width=1.29cm,height=0.98cm]{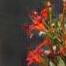}
		\\
		\includegraphics[width=2.7cm,height=1.58cm]{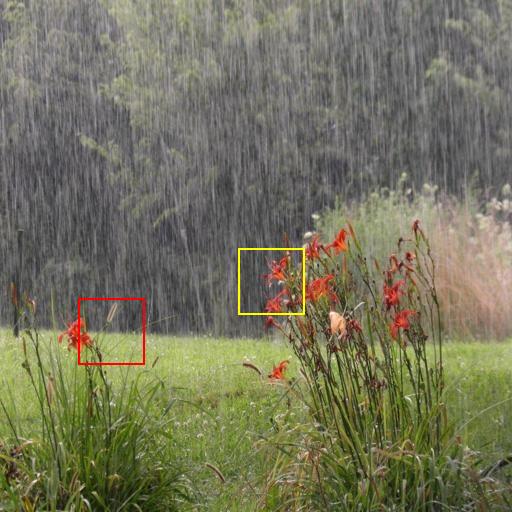}
		\includegraphics[width=2.7cm,height=1.58cm]{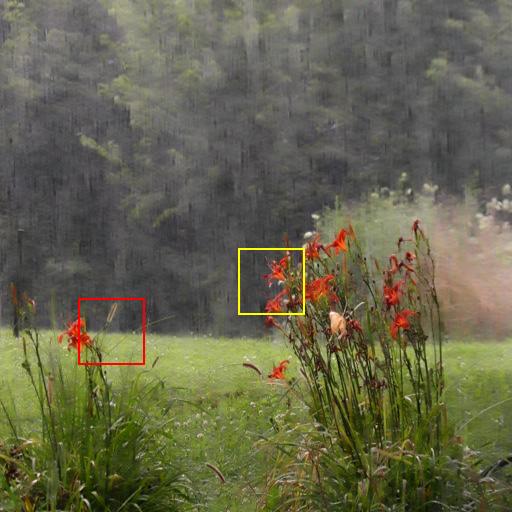}
		\includegraphics[width=2.7cm,height=1.58cm]{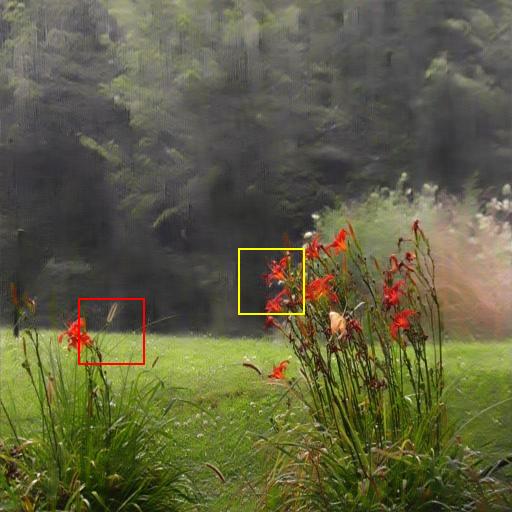}\\
		(d)\hskip70pt(e)\hskip70pt(f)\\
		\includegraphics[width=1.29cm,height=0.98cm]{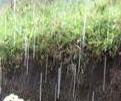}
		\includegraphics[width=1.29cm,height=0.98cm]{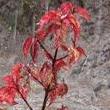}
		\includegraphics[width=1.29cm,height=0.98cm]{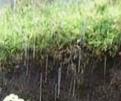}
		\includegraphics[width=1.29cm,height=0.98cm]{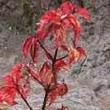}
		\includegraphics[width=1.29cm,height=0.98cm]{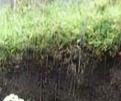}
		\includegraphics[width=1.29cm,height=0.98cm]{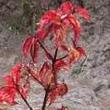}
		\\
		\includegraphics[width=2.7cm,height=1.58cm]{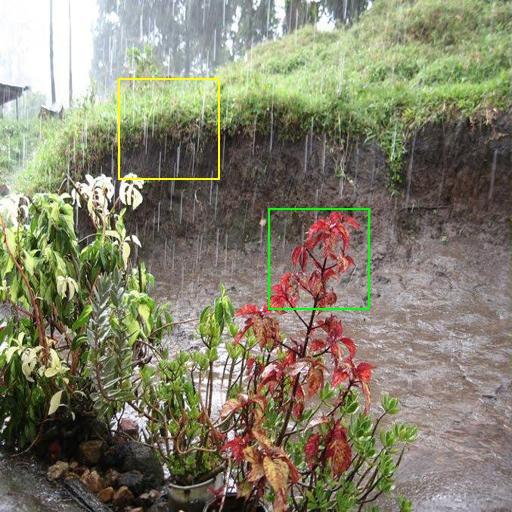}
		\includegraphics[width=2.7cm,height=1.58cm]{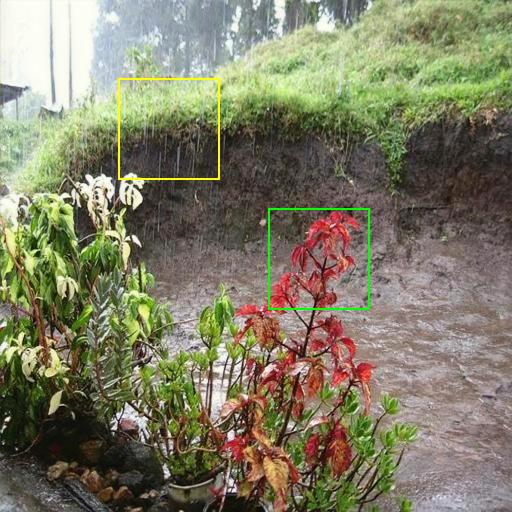}
		\includegraphics[width=2.7cm,height=1.58cm]{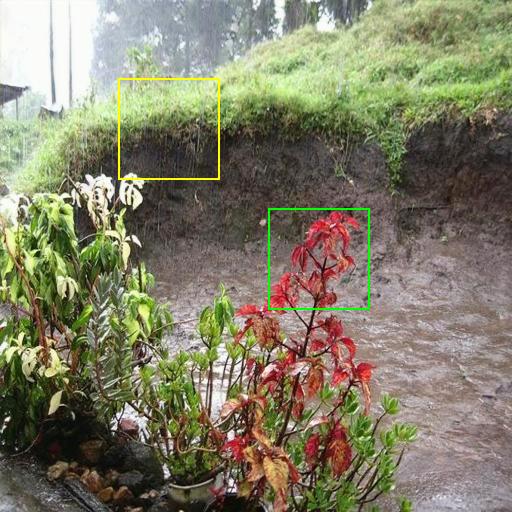}\\
		(g)\hskip70pt(h)\hskip70pt(i)
		\caption{Sample image deraining results.  (a),(d),(g) are Rainy images. (b)  De-rained using DID-MDN \cite{Authors18} where zoomed in parts show the rain streaks and blurry effects on faces and hands. (e) De-rained using Fu et al. \cite{Authors17f} where zoomed in parts show under de-raining of the image. (h) De-rained using JORDER~\cite{Authors17h} where zoomed in  part shows under de-raining near grass and blurry effects on the plant. (c),(f),(i) are de-rained images using QuDeC.}
		\label{Fig:Initial}
		\vskip-20pt
	\end{figure}
	
	In this paper, we address the problem of removing rain streaks from a single rainy image.  Rain streak removal or image de-raining is a difficult problem since a rainy image may contain rain streaks which may vary in size, direction and density.  A number of different techniques have been developed in the literature to address this problem.   These algorithms can be clustered into two main groups - (i) video based algorithms \cite{Authors6,Authors7,Authors15c,Authors18e,Authors18f}, and (ii) single image-based algorithms \cite{Authors18,Authors17f,Authors16, Authors17h, Authors17c}. Algorithms corresponding to the first category assume temporal consistency among the image frames, and use this assumption for de-raining. On the other hand, single image de-raining methods attempt to use some prior information to remove rain components from a single image \cite{Authors16, Authors12,Authors17c,Authors18}.  Priors such as sparsity \cite{Authors17g, Authors15} and low-rank representation \cite{Low_Rank_ICCV2013} have been used in the literature.  In particular, the method proposed by Fu et al. \cite{Authors17d} uses a priori image domain knowledge by focusing on high frequency details during training to improve the de-raining performance.  However, it was shown in \cite{Authors18}, that this method tends to remove some important parts  in  the  de-rained  image (see Figure~\ref{Fig:Initial}(e)).   Similarly, a recent work by Zhang  and Patel \cite{Authors18} uses the image-level priors to estimate the rain density information which is then used for de-raining.  Although their approach provides the state-of-the-art results, they estimate image level priors which do not consider the location information of rain drops in the image.  As a result, their algorithm tends to introduce some artifacts in the final de-rained images.  These artifacts can be clearly seen from the de-rained results shown in Figure~\ref{Fig:Initial}(b).  
	
\begin{figure}[htp!]
    	\centering
    	\includegraphics[width=2.7cm,height = 1.58cm]{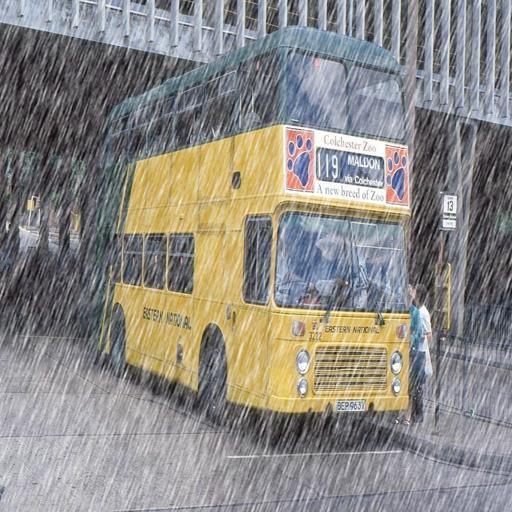}
    	\includegraphics[width=2.7cm,height = 1.58cm]{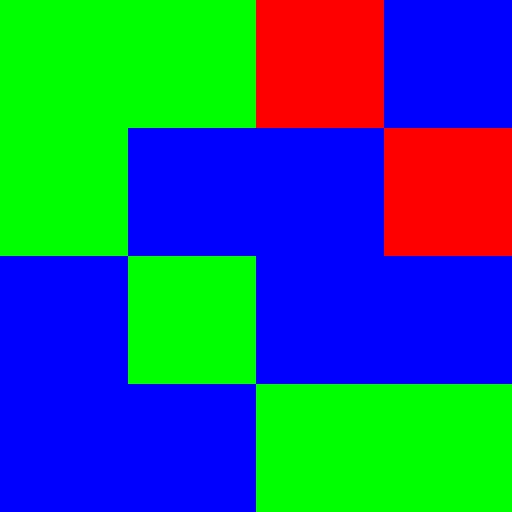}
    	\includegraphics[width=2.7cm,height = 1.58cm]{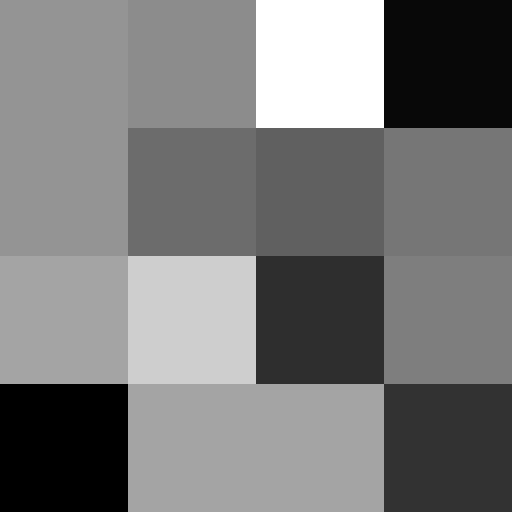}\\
    	(a)\hskip60pt(b)\hskip60pt(c)\\
    	\caption{(a) Input rainy image, $y$.  (b) location-quality-label-map. (c) corresponding confidence scores. }
    	\label{Fig:exp_sg}
    \end{figure}

	In our previous work \cite{Authors19} we proposed a different approach to image de-raining where we made use of the observation that rain streak density and direction does not change drastically with different scales. Rather than relying on the rain density information (i.e. heavy, medium or light) alone, we developed a method in which the rain streak location information is also taken into consideration in a multi-scale fashion  to improve the de-raining performance.   In this paper, we extend our work in \cite{Authors19} by  also incorporating the distortion level information at each location in the lower scales of the image to further improve the de-raining performance.  To this end, we use the Natural Image Quality Evaluator (NIQE) no-reference image quality score~\cite{mittal2012making} and formulate a joint task of estimating the level of distortion at each spatial location of the image and computing rain streak information to reconstruct the clean image. This is achieved by adding a decoder network which computes the level of distortion at each location, to our  UMRL approach in \cite{Authors19}.  This additional decoder network labels the distortion at each location into three classes and also computes the confidence scores for each patch (which indicates how confident the decoder is about those obtained labels) as shown in the Figure~\ref{Fig:exp_sg}. As a result, this can be interpreted as soft labelling of the distortion levels in the image.  As shown in Figure~\ref{Fig:exp_sg} different patches in image are labelled with different confident scores.   This makes the proposed  QuDeC network more robust by blocking the errors in level of distortion while estimating the de-rained image. In addition, we propose a novel loss function to train the network.    In summary, we add the following novelties to our previous UMRL approach in~\cite{Authors19}.
	\begin{itemize}
		\item We formulate a joint task of estimating the level of distortion at each spatial location of the image and compute rain streak information by adding a new decoder to UMRL~\cite{Authors19}.
		\item A new loss function is proposed for estimating the confidence scores for both residual maps and distortion quality-label maps.
	\end{itemize}
	Fig. \ref{Fig:Initial}(c), (f) and (i) present sample results from our QuDeC network, where one can clearly see that  QuDeC is able to remove the noise artifacts and provides better results as compared to \cite{Authors17f,Authors17h} and \cite{Authors18}. To  summarize,  the  following  are  our  key  contributions  in this paper.
	\begin{itemize}
		\item A novel method called QuDeC is proposed which jointly computes the level of distortion at each location and estimates rain streak information in multi-scale fashion to obtain the de-rained image.
		\item Confidence scores are computed for each task and judiciously combined to obtain improved results. 
		\item We run extensive experiments to show the performance of QuDeC against several recent state-of the-art approaches on both synthetic and real rainy images. Furthermore, an ablation study is conducted to demonstrate the effectiveness of different parts of the proposed QuDeC network.
	\end{itemize}
	
	Rest  of  the  paper  is  organized  as  follows.  In  Section~\ref{sec:related},  we review  a  few  related  works.  Details  of  the  proposed  method are given in Section~\ref{sec:proposed}. Training details are provided in Section~\ref{sec:training}.  Experimental results are presented in Section~\ref{sec:results}, and finally, Section~\ref{sec:conclusion} concludes the paper with a brief summary. 
	
	\section{Background and Related Work}\label{sec:related} 
	An observed rainy image $y$ can be modeled as the superposition of a rain component (i.e. residual map) $r$ with a clean image $x$ as follows
	\setlength{\belowdisplayskip}{0pt} \setlength{\belowdisplayshortskip}{0pt}
	\setlength{\abovedisplayskip}{0pt} \setlength{\abovedisplayshortskip}{0pt}
	\begin{equation}
	y = x + r.
	\label{eq:formulation}
	\end{equation}
	Given $ y $ the goal of image de-raining is to estimate $x$.  This can be done either by estimating the residual map $r$ and then subtracting it from the observed image $y$ or by directly estimating $x$ from $y$.  Various methods have been proposed in the literature for image de-raining \cite{Authors12,Authors14d,Authors15,Authors14b,Authors16} including dictionary learning-based \cite{Authors9}, Gaussian mixture-model (GMM) based \cite{Authors2000}, and low-rank representation based \cite{Authors13} methods.   In recent years, deep learning-based single image de-raining methods have also been proposed in the literature.  Fu et al. \cite{Authors17d} proposed a convolutional neural network (CNN) based approach in which they directly learn the  mapping  relationship  between  rainy  and  clean  image  detail layers  from  data. Zhang et al. \cite{Authors17e} proposed a generative adversarial network (GAN) based method for image de-raining.  Furthermore, to  minimize  the artifacts  introduced  by  GANs and  ensure  better  visual  quality,  a  new  refined  loss  function was also  introduced in \cite{Authors17e}.  Fu et al. \cite{Authors17f}  presented an end-to-end deep learning framework for removing rain from individual images using a deep detail network which directly reduces the mapping range from input to output.  Zhang and Patel \cite{Authors18}  proposed  a density-aware multi-stream densely connected CNN for joint rain density  estimation  and  de-raining.  Their  network  automatically  determines  the rain-density  information and  then  efficiently  removes  the corresponding  rain-streaks  using the estimated  rain-density label.  Note the methods proposed in \cite{Authors17f}, and \cite{Authors18} showed the benefits of using multi-scale networks for image de-raining.   Recently, Wang et al. \cite{Authors17b} proposed a hierarchical approach based on estimating different frequency details of an image to get the de-rained image.   The method proposed by Qian et al. \cite{Authors18b} generates attentive maps using the recurrent neural networks, and then uses the features from different scales to compute the loss for removing the rain drops on glasses.   Note that this method was specifically designed for removing rain drops from a glass rather than removing rain streaks from an image. \cite{NonLocal2018,Xu_2018_CVPR,Authors18d} illustrated the importance of attention based methods in low-level vision tasks. In a recent work \cite{Authors18d}, Li et al. proposed a convolutional and recurrent neural network-based method for single image de-raining which makes use of the contextual information for rain removal.   It was observed in \cite{Authors18}, that some of the recent deep learning-based methods tend to under de-rain or over de-rain  the image if the rain condition present in the rainy image is not properly considered during training.
	
	\begin{figure*}[htp!]
		\begin{center}
			\centering
			\includegraphics[width=2.7cm,height=1.58cm]{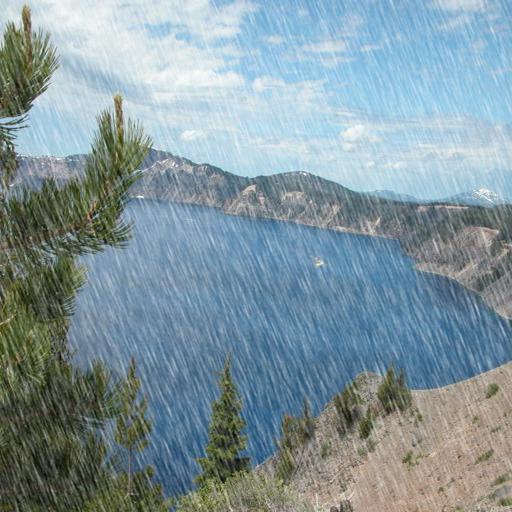}
			\includegraphics[width=2.7cm,height=1.58cm]{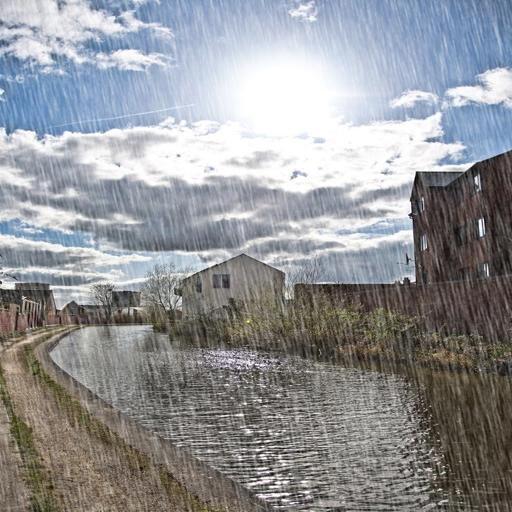}
			\includegraphics[width=2.7cm,height=1.58cm]{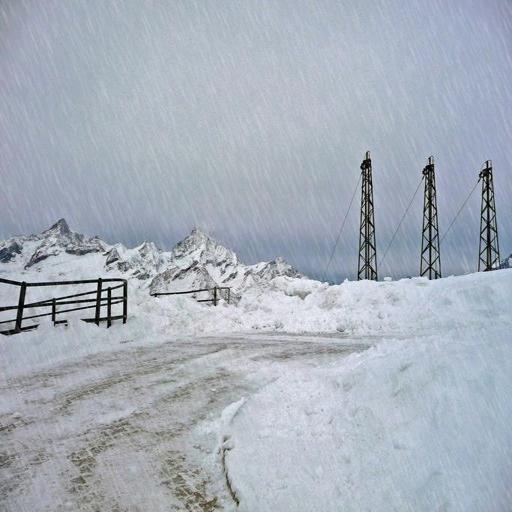}
			\includegraphics[width=2.7cm,height=1.58cm]{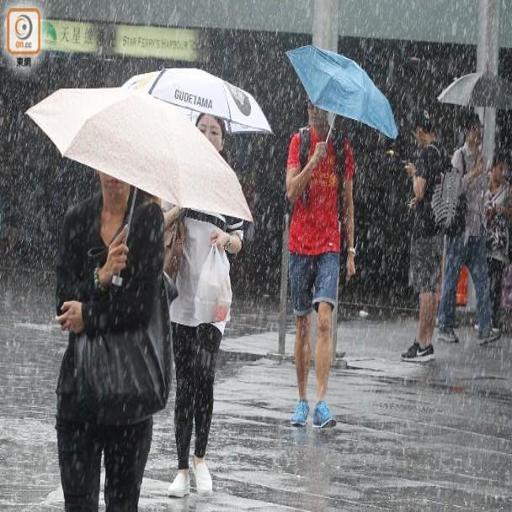}
			\includegraphics[width=2.7cm,height=1.58cm]{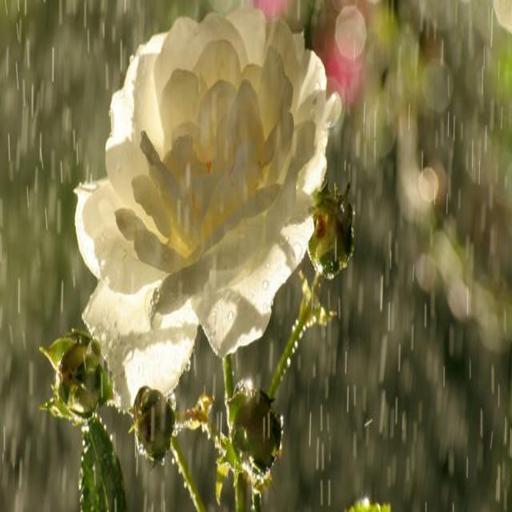}
			\includegraphics[width=2.7cm,height=1.58cm]{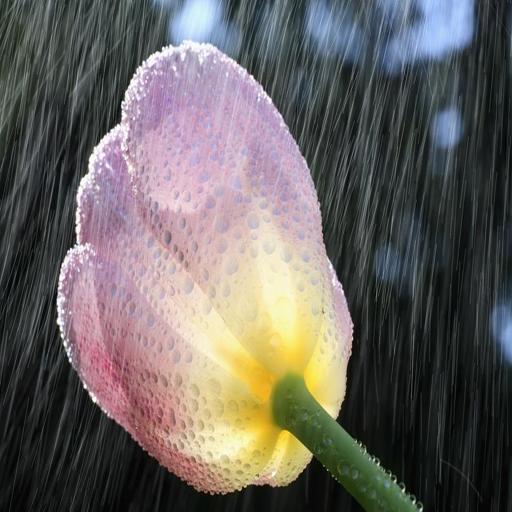}\\
			\includegraphics[width=2.7cm,height=1.58cm]{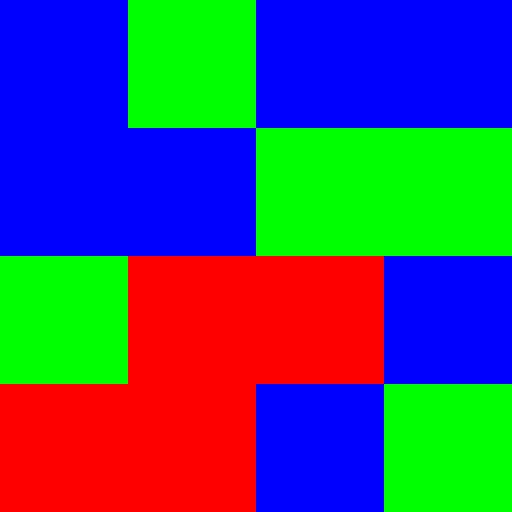}
			\includegraphics[width=2.7cm,height=1.58cm]{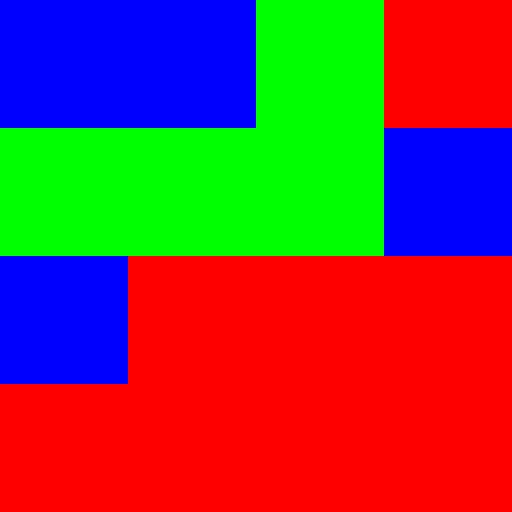}
			\includegraphics[width=2.7cm,height=1.58cm]{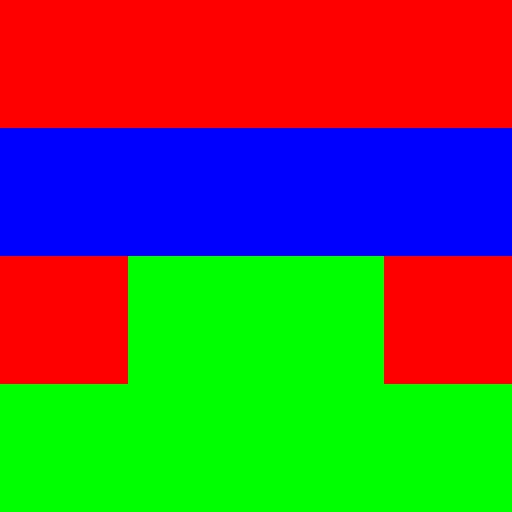}
			\includegraphics[width=2.7cm,height=1.58cm]{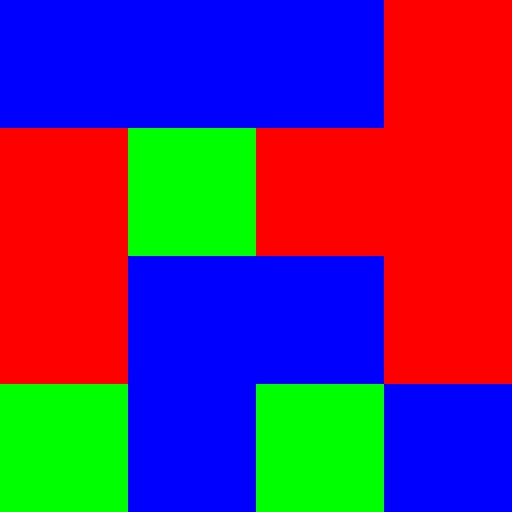}
			\includegraphics[width=2.7cm,height=1.58cm]{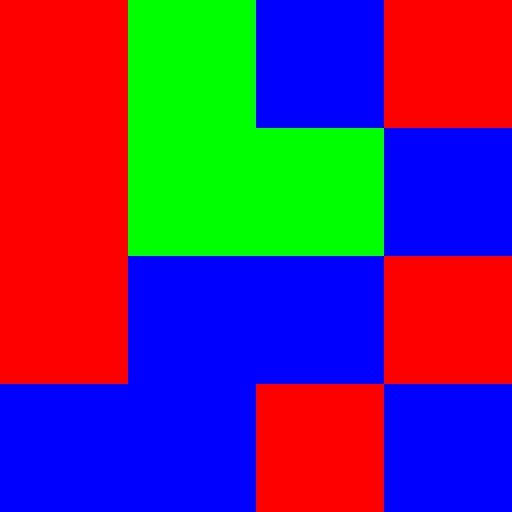}
			\includegraphics[width=2.7cm,height=1.58cm]{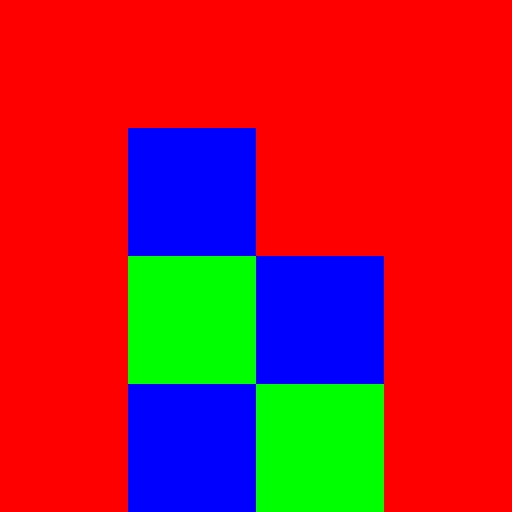}
			\caption{Sample location-quality-label-maps $s$ corresponding to synthetic and real-world rainy images.  Here $red\:color$ indicates high distortion, $blue\:color$ indicates medium distortion, $green\:color$ indicates low distortion in the corresponding patch of the rainy image, $y$.}
			\label{Fig:Sample_segs}
		\end{center}
		\vskip-15pt
	\end{figure*}
	
	\section{Proposed Method}\label{sec:proposed}
	Unlike many deep learning-based methods that directly estimate the de-rained image from the noisy observation, we take a different approach where we formulate the rain removal problem as the joint task of computing distortion level at each location and estimating the residual rain streak component  $\hat{r}$ (i.e residual map). We compute the distortion level for the patches of size $128 \times 128$, and classify them into three classes, $green - low~distortion$, $ blue - medium~distortion$, and $ red - high~distortion$. Note that, the ground truth labels for these patches are computed using the NIQE measure~\cite{mittal2012making}. In this way, the distortion level at each location in the rainy image is represented by the location-quality-label-map $s$. Representing the distortion level at each location, and giving it as a prior information while reconstructing the image helps the network to learn the rain streak information which may vary spatially or affect the background differently.  From Figure ~\ref{Fig:Sample_segs} we can clearly observe that different levels of rain streak affect the background differently.  In some cases the same level of rain streaks have different affects on the background. For example, as can be seen from  Figure~\ref{Fig:Sample_segs}, rain steaks produce different distortion in bright and heavily textured regions.  On the other hand the distortion level is relatively high on the  dark and homogeneous regions.   To handle these regions differently, we propose a method that jointly estimate the level of distortion at each location and uses it as a prior information in estimating the rain streak information (i.e residual map).  In contrast to other methods \cite{Authors18,Authors17h} that use either rain streak binary masks or level of rain streak present in the image, we estimate the distortion caused by rain streaks at each location based on the NIQE measure which makes our method more general and applicable for synthetic and well as real-world images. 
	
\begin{figure}[htp!]
		\centering
		\includegraphics[width=0.48\textwidth,height=5.5cm]{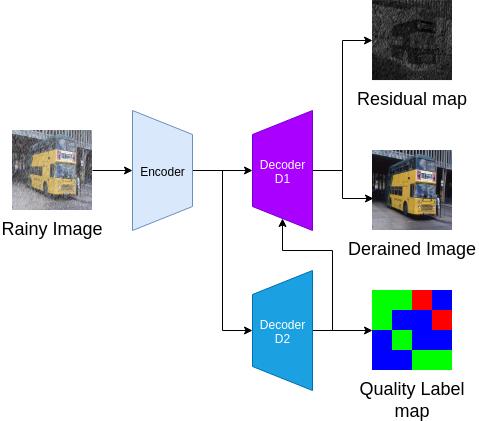}\\
		\caption{An overview of the QuDeC network.}
		\label{Fig:QuDeC_o}
	\end{figure}

	To perform the joint task of computing the location quality label maps ($s$), and estimating the residual maps $\hat{r}$, QuDeC is constructed using one encoder and two decoders as shown in Figure~\ref{Fig:QuDeC_o}. A rainy image is first passed through an encoder network to obtain intermediate features.  These features are then further processed by the decoders.  In particular, Decoder D2 estimates the location quality labels, which are then given as a prior information to Decoder D1 to estimate the residual maps ($r$).  
	
	\subsection{QuDeC Network}
	Given a rainy image $y$, we estimate the residual map and its corresponding confidence maps at three different scales, $\:\{\hat{r}_{\x1},c_{\x1}\} $ (at the original input size), $\:\{\hat{r}_{\x2},c_{\x2}\} $ (at 0.5 scale of the input size), and $\:\{\hat{r}_{\x4},c_{\x4}\} $ (at 0.25 scale of the input size) in a multi-scale fashion \cite{Authors19}, as well as the location-quality-label-map $\hat{s}$ along with the corresponding confidence scores ($c_s$) for label map $s$.
	
\begin{figure*}[htp!]
		\begin{center}
			\centering
			\includegraphics[width=12cm]{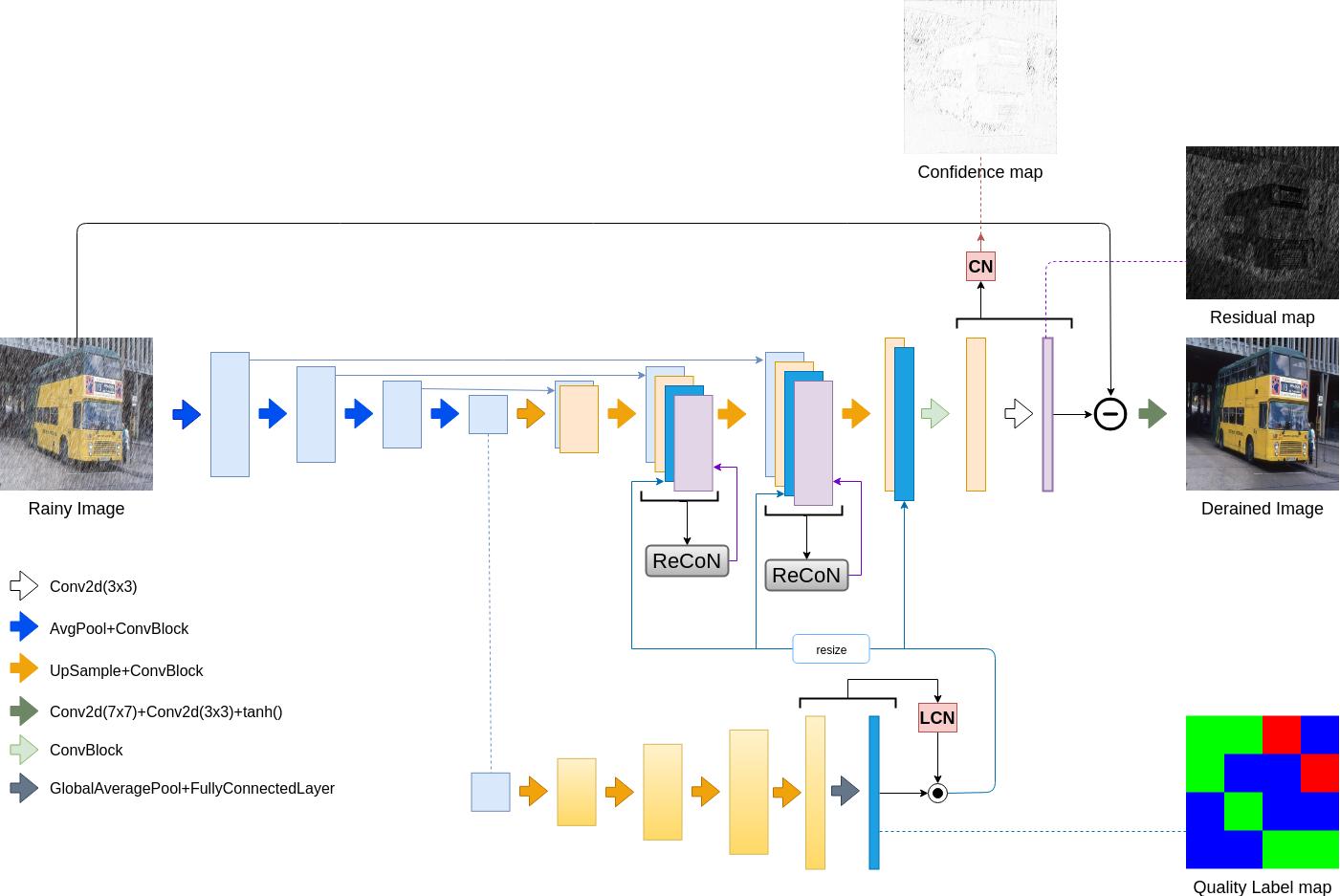}\\
			\caption{An overview of the proposed QuDeC network.  The aim of the QuDeC network is to estimate the clean image given the corresponding rainy image. QuDeC learns the residual maps, location-quality-label-maps, and computes the confidence maps to guide the network. To achieve this, we introduce ReCoN and LCN networks, and feed their outputs to the subsequent layers of Decoder D1.}
			\label{Fig:QuDeC}
		\end{center}
	\end{figure*}

	The QuDeC network aims to judiciously combine the pixel-wise rain residual information from lower scales, and patch-wise distortion level in the quality label maps while estimating the final de-rained image. To achieve this we compute the confidence maps for the estimated residual maps and location-quality-label-maps.  These confidence maps block the errors in the computation by giving low confidence values to erroneous regions in the computations of $\hat{r}$ and $\hat{s}$. Additionally, these confidence scores enable QuDeC to perform soft labeling of the distortion levels in the location-quality-label-map $s$, which makes QuDeC more robust to location quality label errors and effectively use this information as a prior while estimating $\hat{r}$. To compute $\hat{r}$ and $\hat{s}$ we start with UMRL \cite{Authors19} and add Decoder D2 to obtain the network architecture of QuDeC as shown in Figure~\ref{Fig:QuDeC}. 
	
	\begin{figure}[htp!]
		\centering
		\includegraphics[width=2.7cm,height = 1.58cm]{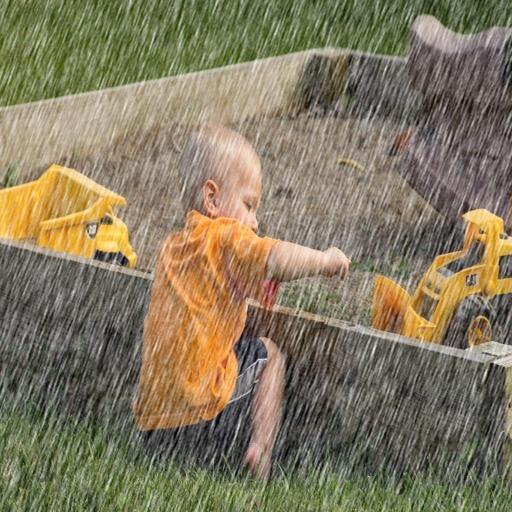}
		\includegraphics[width=2.7cm,height = 1.58cm]{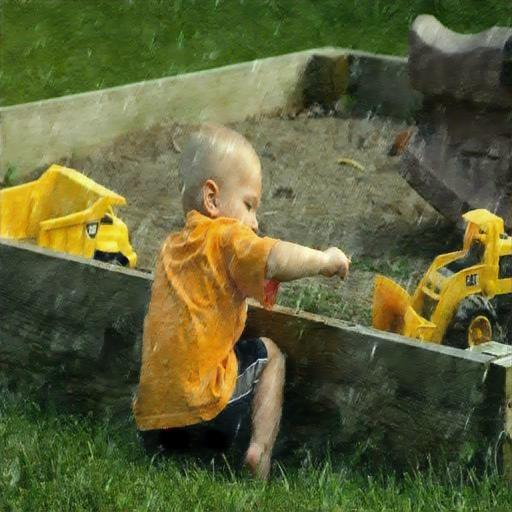}
		\includegraphics[width=2.7cm,height = 1.58cm]{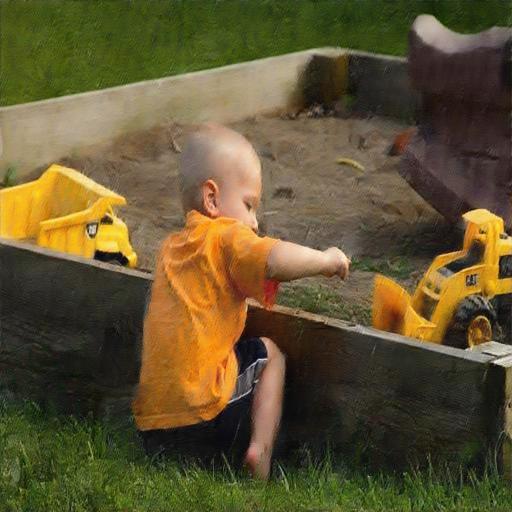}\\
		(a)\hskip60pt(b)\hskip60pt(c)\\
		\includegraphics[width=2.7cm,height = 1.58cm]{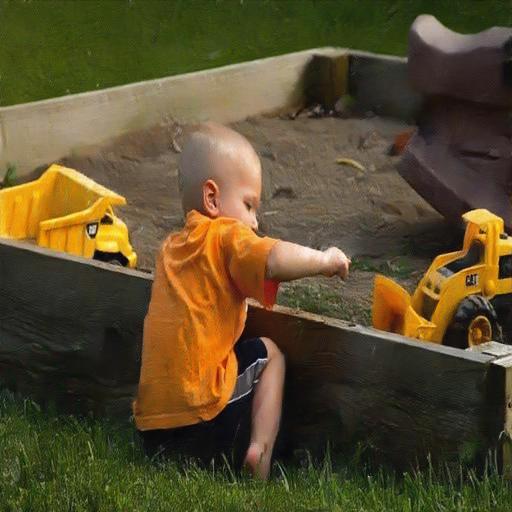}
		\includegraphics[width=2.7cm,height = 1.58cm]{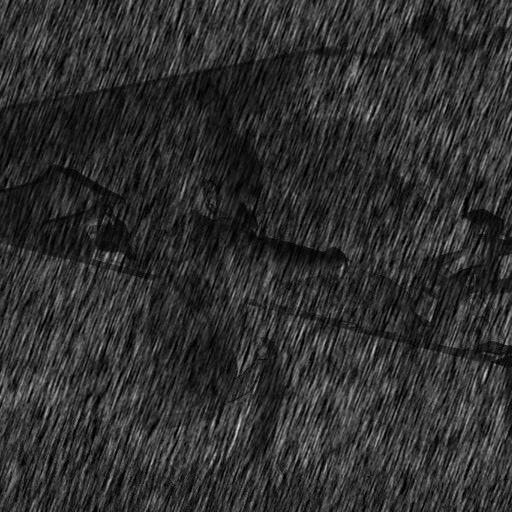}
		\includegraphics[width=2.7cm,height = 1.58cm]{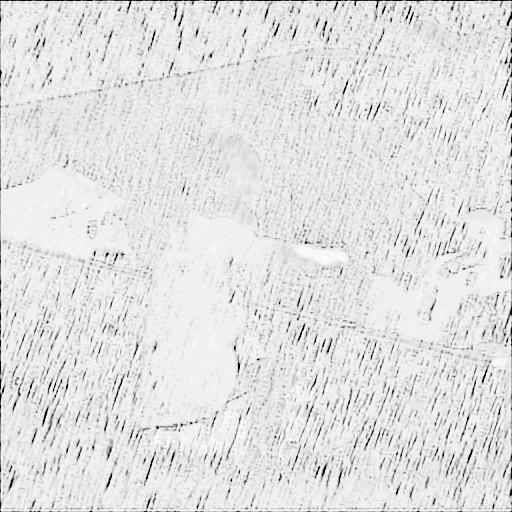}\\
		(d)\hskip60pt(e)\hskip60pt(f)\\
		\caption{(a) Input rainy image, $y$.  (b) De-rained image using the base network. (c) De-rained using \cite{Authors18}. (d) De-rained using the proposed encoder and decoder D1 of the QuDeC network.  (e) The residual map.  (f) The confidence map at scale 1.0($\x1$).}
		\label{Fig:exp1}
	\end{figure} 
	
	\subsubsection{Encoder and Decoder D1}
	Rain streaks are high frequency components and existing de-raining methods either tend to remove high frequencies that are not rain streaks or do not remove the rain near high frequency components of the clean image like edges as shown in the Figure \ref{Fig:exp1}. To address this issue, one can use the information about the location in the image where the network might go wrong in estimating the residual value.  This can be done by estimating a confidence value corresponding to the estimated residual value and guide the network to remove  the artifacts, especially near the edges. For example, we can observe clearly from Figure \ref{Fig:exp1} that the residual map and its corresponding confidence map are able to capture the regions where there is high probability of incorrect estimates. In the encoder and decoder D1 networks we estimate the residual values and their corresponding confidence maps at different scales (1.0($\x1$), 0.5($\x2$) and 0.25($\x4$)) of the input size.  This information is then fed back to the subsequent layers so that the network can learn the residual value at each location, given the computed residual value and confidence value at lower scales.
	
\begin{figure}[htp!]
		\centering
		\includegraphics[width=1.7cm, height=4.5cm]{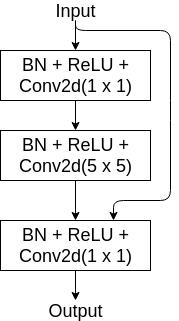}\hskip10pt
		\includegraphics[width=1.7cm, height=4.5cm]{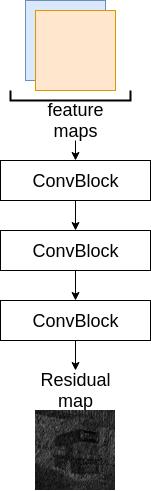}\hskip10pt
		\includegraphics[width=1.7cm, height=4.5cm]{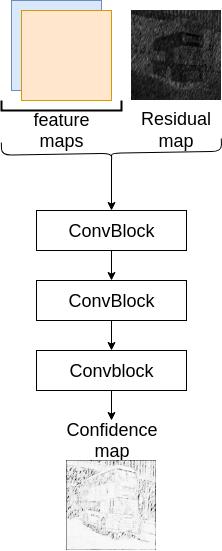}\hskip10pt
		\includegraphics[width=2.25cm, height=4cm]{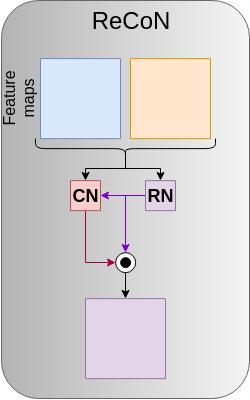}\\
		(a)\hskip50pt(b)\hskip45pt(c)\hskip50pt(d)\\
		\caption{(a) Convolutional block (ConvBlock).  BN - batchnormalization, ReLU - Rectified Linear Units, Conv2d($m\times m$) - convolutional layer with kernel of size $m\times m$. (b) Residual Network (RN). (c) Confidence map Network (CN). (d) Residual-Confindence Network(ReCoN).}
		\label{Fig:blocks}
	\end{figure}

	The encoder and decoder D1 networks are similar to the the encoder and decoder networks of UMRL \cite{Authors19} where a convolutional block (ConvBlock as shown in Figure~\ref{Fig:blocks}(a)) is used as the building block. The encoder network is described as follows,\\
	ConvBlock(3,32)-AvgPool-ConvBlock(32,32)-AvgPool-Convblock(32,32)-AvgPool-ConvBlock(32,32)-AvgPool\\
	where AvgPool is the average pooling layer, UpSample is the upsampling convolution layer, and ConvBlock$(i, j)$ indicates ConvBlock with $i$ input channels and $j$ output channels. The decoder D1 network is described as follows,\\
	ConvBlock(32,32)-UpSample-ConvBlock(64,32)-UpSample-ConvBlock(67,32)-UpSample-ConvBlock(67,16)-ConvBlock(16,16)-Conv2d$(3\times 3)$,\\
	where ReCoN networks are added to Decoder D1 to estimate $\hat{r}$ at different scales and their corresponding confidence maps $c$. Given the feature maps as input to ReCoN network, RN (residual network) estimates the residual map $\hat{r}$ and CN (confidence network) computes the corresponding confidence map $c$ as shown in  Figure~\ref{Fig:blocks}(d).
	
	Feature maps at different scales such as  $\x2$ and $\x4$ are given as inputs to the Residual Network (RN) to estimate the residual map at the corresponding scale as shown in  Figure~\ref{Fig:blocks}(d). RN consists of the  following sequence of convolutional layers,
	\noindent  Convblock(64,32)-Convblock(32,32)-Convblock(32,3)\\
	\noindent   as shown in Figure \ref{Fig:blocks}(b). We use the estimated residual map and the feature maps as inputs to the Confidence map Network (CN) to compute the confidence measure at every pixel, which indicates how sure the network is about the residual value at each pixel. CN consists of the following sequence of convolutional layers,\\
	
	\noindent Convblock(67,16)-Convblock(16,16)-Convblock(16,3)
	
	\noindent as shown in Figure \ref{Fig:blocks}(c). Given the estimated residual map and the corresponding feature maps as inputs to the confidence map network, it estimates $c_{\x4}$ and $c_{\x2}$. The element wise product of $\hat{r_i}$ and $c_i$ is computed, and up-sampled.  This is used an input to the subsequent layers of the Decoder D1 network in QuDeC as shown in Figure \ref{Fig:QuDeC} for $i\in \{\x2,\x4\}$. Given the output residual map $r_{\x1}$ and the feature maps of the final layer of the Decoder D1 network in QuDeC as input to CN, we get $c_{\x1}$.
	
	A Refinement Network (RFN) is used at the end of Decoder D1 to produce the de-rained image. It takes $y_i - \hat{r}_i$ as the input and generates $\hat{x}_i$ (i.e. derained image) as the output.  The RFN consists of the following blocks\\
	Conv2d$(7 \times 7)$-Conv2d$(3 \times 3)$-tanh(),
	where Conv2d$(m \times m)$ represents 2D convolution using the kernel of size $m \times m$.
	
	\begin{figure}[htp!]
		\centering
		\begin{minipage}{.155\textwidth}
			\centering
			\caption*{\emph{\footnotesize{PSNR:22.29 SSIM: 0.86}}}
			\includegraphics[width=2.7cm,height = 1.8cm]{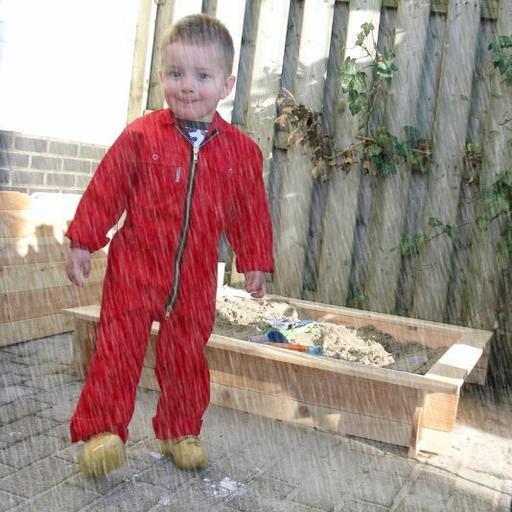}
			\captionsetup{labelformat=empty}
			\captionsetup{justification=centering}
		\end{minipage}
		\begin{minipage}{.155\textwidth}
			\centering
			\caption*{\emph{\footnotesize{PSNR:26.63 SSIM: 0.91}}}
			\includegraphics[width=2.7cm,height = 1.8cm]{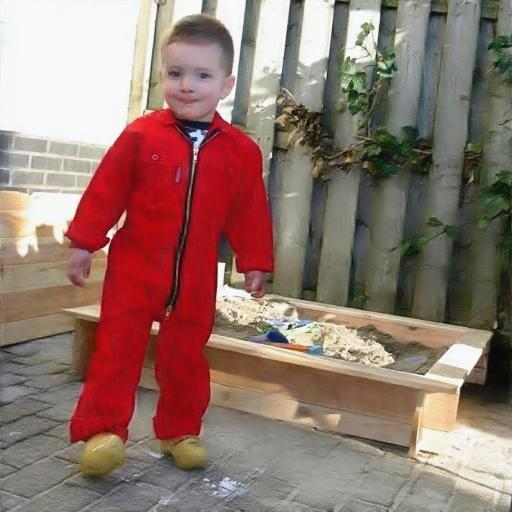}
			\captionsetup{labelformat=empty}
			\captionsetup{justification=centering}
		\end{minipage}
		\begin{minipage}{.155\textwidth}
			\centering
			\caption*{\emph{\footnotesize{PSNR:30.39 SSIM: 0.94}}}
			\includegraphics[width=2.7cm,height = 1.8cm]{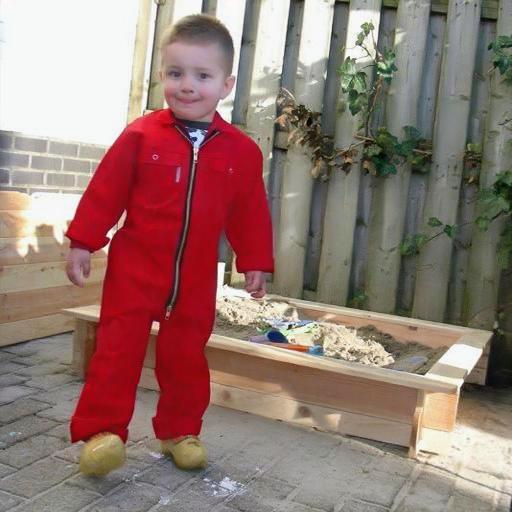}
			\captionsetup{labelformat=empty}
			\captionsetup{justification=centering}
		\end{minipage}\\ \vskip2pt
		(a)\hskip70pt(b)\hskip70pt(c)\\
		\begin{minipage}{.155\textwidth}
			\centering
			\caption*{\emph{\footnotesize{PSNR:33.84 SSIM: 0.97}}}
			\includegraphics[width=2.7cm,height = 1.8cm]{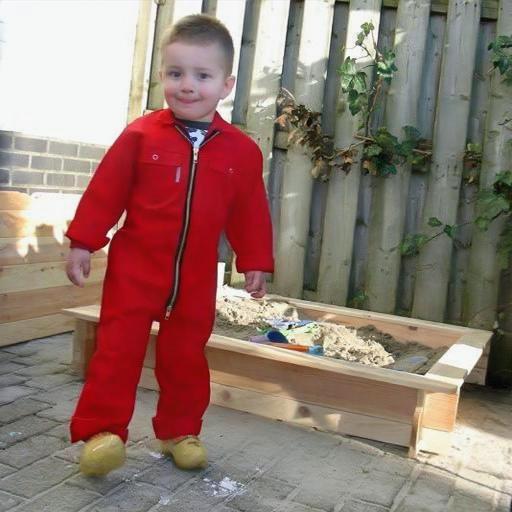}
			\captionsetup{labelformat=empty}
			\captionsetup{justification=centering}
		\end{minipage}
		\begin{minipage}{.155\textwidth}
			\centering
			\caption*{\emph{\footnotesize{       }}}
			\includegraphics[width=2.7cm,height = 1.8cm]{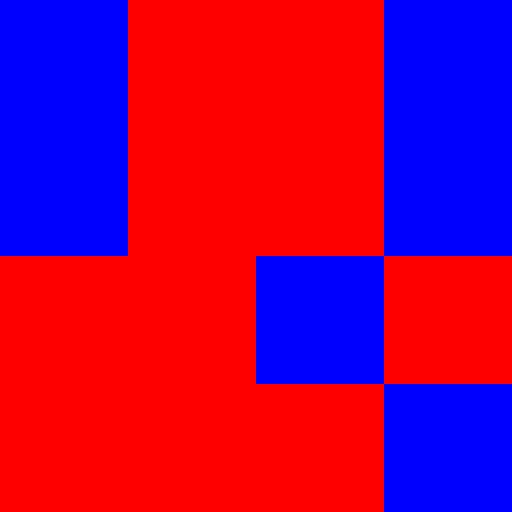}
			\captionsetup{labelformat=empty}
			\captionsetup{justification=centering}
		\end{minipage}
		\begin{minipage}{.155\textwidth}
			\centering
			\caption*{\emph{\footnotesize{PSNR:Inf SSIM: 1}}}
			\includegraphics[width=2.7cm,height = 1.8cm]{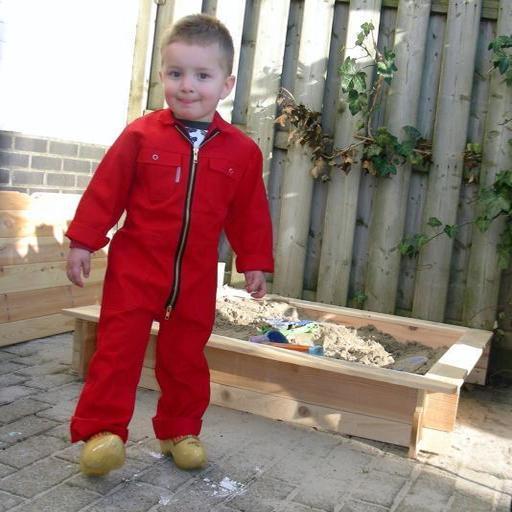}
			\captionsetup{labelformat=empty}
			\captionsetup{justification=centering}
		\end{minipage}\\ \vskip2pt
		(d)\hskip70pt(e)\hskip70pt(f)\\
		\caption{(a) Input rainy image. (b) De-rained image using DIDMDN \cite{Authors18}. (c) De-rained using UMRL \cite{Authors19}. (d) De-rained using our proposed method QuDeC.  (e) Location-quality-label-map $s$ computed by QuDeC.  (f) Ground truth clean image.}
		\label{Fig:exp3}
	\end{figure}

	\subsubsection{Decoder D2}
	Rain streaks have different distortion effects on different parts of the image.  
	
\begin{figure}[h!]
		\centering
		\includegraphics[width=0.48\textwidth]{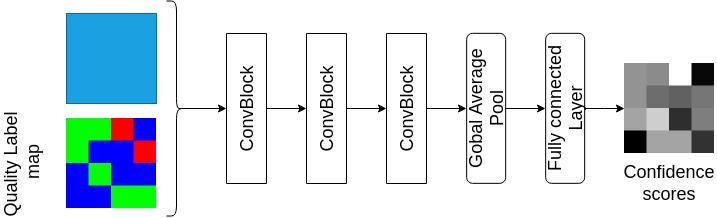}\\
		\caption{Label Confidence Network (LCN).}
		\label{Fig:LCN}
		\vskip-20pt
	\end{figure}

	As a result while reconstructing the clean image, we also estimate the prior information such as distortion caused by rain streak at every location. Not using this prior information may lead to inferior de-raining performance. As can be seen from Figure~\ref{Fig:exp3} that DIDMDN~\cite{Authors18} and UMRL~\cite{Authors19} do not perform well as they lack prior distortion level information at each location of the image. We address this by introducing a decoder D2 in our method and formulate joint task of computing the distortion level at each location and estimating the residual rain streak information. Decoder D2 is similar to Decoder D1 with the following sequence of blocks,\\
	ConvBlock(32,32)-UpSample-ConvBlock(64,32)-UpSample-ConvBlock(67,32)-UpSample-ConvBlock(67,16)-ConvBlock(16,16)-GlobalAveragePool-FullyConnectedLayer.\\
	Decoder D2 takes feature maps obtained from the encoder as shown in Figure~\ref{Fig:QuDeC}, to estimate the distortion levels in location-quality-label-map. Label Confidence Network (LCN) is used to compute the confidence scores corresponding to the location-quality-label-maps $\hat{s}$. Feature maps from the last layer of D2 and obtained location-quality-label-maps $\hat{s}$ are fed to LCN to compute the confidence scores $c_s$ as shown in  Figure~\ref{Fig:LCN}. LCN is a sequence of three ConvBlocks followed by global average pool and fully connected layers as shown in Figure~\ref{Fig:LCN}.
	
	\subsection{Loss for QuDeC Network}
	In image restoration tasks, maximum-a-posteriori method is often used to optimize the network parameters ($\theta$)  as follows,
	\begin{equation}
	\hat{\theta} = \argmax_{\theta} P(f_{\theta}(y)|y;\theta) =  \argmax_{\theta} P(\hat{x}|y;\theta),
	\label{Eq:eq1}
	\end{equation}
	where $P(.)$ is the probability function and $f_{\theta}(.)$  represents the QuDeC network, $\hat{x} = f_{\theta}(y)$. Since QuDeC performs the joint task of estimating $\hat{r}$ and $\hat{s}$, the above optimization can be updated as follows,
	\begin{equation}
	\nonumber
	\hat{\theta} =  \argmax_{\theta} P(\hat{x},\hat{s}|y;\theta) = \argmax_{\theta} P(\hat{x}|y,\hat{s};\theta)P(\hat{s}|y;\theta).
	\label{Eq:eq2}
	\end{equation}
	To find the optimal network parameters $\hat{\theta}$,  $P(\hat{x},\hat{s}|y;\theta)$ needs to be maximized. For simplicity to solve this optimization problem, let us assume $P(\hat{x}|y,\hat{s};\theta)$ and $P(\hat{s}|y;\theta)$ are Gaussian distributions. As our goal is to minimize the joint errors between $\hat{x}$ and the ground-truth clean image ($x$), $\hat{s}$ and the ground-truth labels ($s$) of rainy image $y$, we denote the mean of distribution $P(\hat{x}|y,\hat{s};\theta)$ as $x$ and variance as $\sigma^2$, and mean of distribution $P(\hat{s}|y;\theta)$ as $s$ and variance as $\sigma^2_s$. Thus our objective becomes,
	\begin{equation}
	\begin{aligned}
	\hat{\theta} &=  \argmax_{\theta} \log \left( P(\hat{x}|y,\hat{s};\theta)P(\hat{s}|y;\theta) \right) \\
	\hat{\theta} &=  \argmax_{\theta} \log \left( P(\hat{x}|y,\hat{s};\theta)\right) + \log \left(P(\hat{s}|y,\hat{s};\theta) \right) \\
	\hat{\theta} &= \argmax_{\theta} -\frac{1}{\sigma^2}\|\hat{x}-x\|^2_2  -\frac{1}{\sigma^2_s}\|\hat{s}-s\|^2_2 + \log(\frac{1}{\sigma^2 \sigma^2_s}).
	\label{Eq:eq3}
	\end{aligned}
	\end{equation}
	Substituting, $\mathcal{L}_r = \frac{1}{\sigma^2}\|\hat{x}-x\|^2_2$ and $\mathcal{L}_s = \frac{1}{\sigma^2_s}\|\hat{s}-s||^2_2$ into the above equation, we get 
	\begin{equation}
	\begin{aligned}
	\hat{\theta} &= \argmax_{\theta} -\mathcal{L}_r  -\mathcal{L}_s + \log(\frac{1}{\sigma^2}) + \log(\frac{1}{\sigma^2_s})\\
	\hat{\theta} &= \argmax_{\theta} - \mathcal{L}_r + \log(\frac{1}{\sigma^2})  -\left(\mathcal{L}_s - \log(\frac{1}{\sigma^2_s})\right)\\
	\hat{\theta} &= \argmax_{\theta} - \mathcal{L}_r + \mathcal{L}_c -\mathcal{L}_{cs},
	\label{Eq:eq4}
	\end{aligned}
	\end{equation}
	where $\mathcal{L}_c = \log(\frac{1}{\sigma^2})$, and $\mathcal{L}_{cs} = \mathcal{L}_s - \log(\frac{1}{\sigma^2_s})$.
	In \eqref{Eq:eq4} variance ($\sigma^2$) and $\sigma^2_s$ can be inferred in two ways as explained in \cite{KendallGal2017UncertaintiesB,kendall2017multi}. (i) Epistemic uncertainty, which is explained as the model uncertainty given enough data to train, and (ii) Aleatoric uncertainty that captures noise inherent in the observations, which is data dependent. Epistemic uncertainty can be formulated as variational inference to compute variance. Aleatoric uncertainty can be formulated as MAP (maximum-aposterior) or ML (maximum-likelihood) inference. Here, in our method we attempt to address the uncertainty caused in the outputs due to different properties of rain streaks like density, direction, and effect on background scene which are inherent in rainy images.  Following the ML inference, in the above equation \eqref{Eq:eq4}, we can view the terms $\frac{1}{\sigma^2}$ and $\frac{1}{\sigma^2_s}$  as the corresponding location-based confidence maps or scores. These confidence maps  indicate the erroneous regions in the estimates of residual maps $\hat{r}$ or location-quality-label-maps $\hat{s}$ by giving low values to those regions or pixels. These errors occur at regions or pixels where the variance is high.  Note that in our method, the confidence map has no ground-truths. We compute these confidence scores using CN (Confidence Network) and LCN (Label Confidence Network) as explained in the earlier sections. Note that the values in the confidence map at every position will be in the range of $[0,1]$. Additionally  the L2-norm in loss of $\mathcal{L}_s$ in \eqref{Eq:eq4} for classifying the distortion levels in location-quality-map $s$, should be replaced with the cross entropy loss. The residual maps are estimated in a multi-scale fashion. Thus loss is constructed as follows,
	\begin{equation}
	\begin{aligned}
	\mathcal{L}_r &= \sum_{i \in \{\x1,\x2,\x4\}}\lVert(c_i\odot\hat{x}_i)-(c_i\odot x_i)\rVert_2^2, \\
	\mathcal{L}_c &= \sum_{i \in \{\x1,\x2,\x4\}} \big(\sum_j\sum_k \log(c_{i_{jk}})\big),\\
	\mathcal{L}_{cs} &= \sum_{\forall p \in y}c_{s}(p)\mathcal{L}_{CE}(\hat{s}(p),s(p)) -\lambda_{cs} \log(c_{s}(p))\\
	\mathcal{L}_u &= \mathcal{L}_r + \mathcal{L}_{cs} - \lambda_1 \mathcal{L}_c,
	\end{aligned}
	\end{equation}
	where $\mathcal{L}_{CE}(\hat{s}(p),s(p)) = -s(p)\log(\hat{s}(p))-(1-s(p))\log(1-\hat{s}(p))$, and $\{p\}$ is a set of patches in the rainy image $y$. Inspired by the importance of the perceptual loss in many image restoration tasks \cite{Authors16b,Authors17g}, we also use it to further improve the visual quality of the de-rained images. Features from layer $relu1\_2$ of a pretrained network VGG-16 \cite{Authors14c} are used to compute the perceptual loss \cite{Johnson2016Perceptual,zhang2017multistyle}. Let $F(.)$ denote the features obtained using the VGG16 model \cite{Authors14c}, then the perceptual loss is defined as follows
	\begin{equation}
	\mathcal{L}_p = \frac{1}{NHW}\sum_i \sum_j \sum_k \lVert F(\hat{x}_1)^{i,j,k}-F(x_1)^{i,j,k}\rVert_2^{2},
	\end{equation}
	where $N$ is the number of channels of $F(.)$, $H$ is the height and $W$ is the width of feature maps. The overall loss used to train the QuDeC network is,
	\begin{equation}
	\mathcal{L} = \mathcal{L}_r + \mathcal{L}_{cs} - \lambda_1 \mathcal{L}_c + \lambda_2 \mathcal{L}_p.
	\end{equation}

	\section{QuDeC Training}\label{sec:training}
\begin{figure}[htp!]
		\centering
		\includegraphics[width=0.5\textwidth]{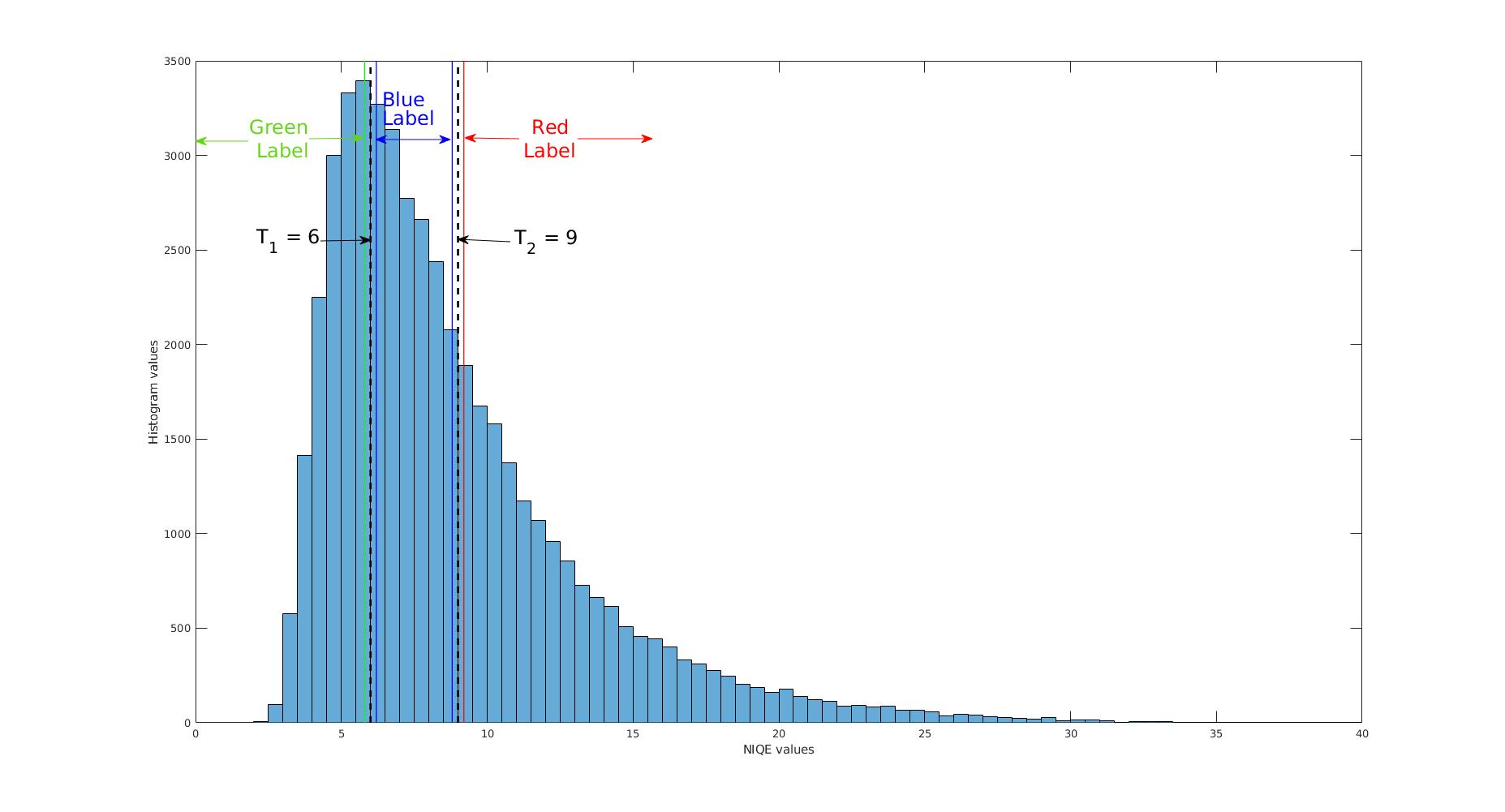}\\
		\caption{Histogram of NIQE values corresponding to the $128\times128$ patches from the DIDMDN training images.}
		\label{Fig:Hist}
		\vskip-20pt
	\end{figure}

	The QuDeC network is trained using the synthetic image datasets created by the authors of \cite{Authors18,Authors17e,Authors17h}. The dataset in  \cite{Authors18} consists of 12000 images with different rain levels like low, medium and high.  The dataset in \cite{Authors17e} contains 700 training images. The dataset in \cite{Authors17h} contains 1800 rainy images for training. We generate the ground truth location-quality-maps, $s$ which indicate the distortion levels of the background in rainy images, using the NIQE scores. Figure~\ref{Fig:Hist} show the histogram of NIQE scores corresponding to the $128\times128$ patches of rainy images from the  DIDMDN training dataset \cite{Authors18}. We divide the patches into three levels of distortions using the thresholds $T_1=6$ and $T_2=9$. These thresholds are chosen such that the patches  are divided into three equal groups.  The following pseudo code summarizes the procedure used for generating the location-quality-label-maps using the NIQE scores.

	\begin{algorithmic}\label{alg:s}
		\STATE Input: Rainy image $y$, patch in $y$ of size $128 \times 128$. 
		\STATE Output: Location-quality-label-map, $s^{\dagger}$.
		\FOR{$\forall p \in y$}
		\IF{NIQE($y(p)) \leq T_1-0.2$}
		\STATE $s^{\dagger}(p) = green$
		\ELSIF{$T_1 + 0.2 <$ NIQE($y(p)) < T_2 -0.2$} 
		\STATE $s^{\dagger}(p) = blue$
		\ELSIF{ NIQE($y(p)) \geq T_2 +0.2$}
		\STATE $s^{\dagger}(p) = red$
		\ENDIF
		\ENDFOR
	\end{algorithmic}
	
	Note that the NIQE scores \cite{mittal2012making} are obtained using various features which may not be specifically useful for computing the distortion levels caused by the rain streaks.  Furthermore, the thresholds $T_1$ and $T_2$ may vary from one dataset to another.  In order to deal with these issues, we train a Generation-of-Labels Network (GLN) to automatically generate the location-quality-label-maps $s$ from the input rainy images. The GLN network is trained using pairs of $(y,s^{\dagger})$ from the DIDMDN training images. We provide the network architecture and training details of GLN in the appendix. The estimated  $s$ from GLN as well as the clean image $x$ along with the rainy image $y$ are used to train the proposed QuDeC network.

	\subsection{Test Datasets}
	The QuDeC network is tested on synthetic and real-world rainy images published by the authors of \cite{Authors18,Authors17e,Authors17h}. The DIDMDN synthetic test dataset consists of two subsets \textit{Test-1} and \textit{Test-2} containing 1000 and 1200 images, respectively \cite{Authors18}.  The \textit{Rain800} dataset shared by the authors of \cite{Authors17e} contains 100 synthetic rainy images for testing. The \textit{Rain200H} dataset contains 200 heavy rain synthetic images provided by the authors \cite{Authors17h}. In addition to the synthetic images, we use 100 real-world rainy images provided by the authors of \cite{Authors18,Authors17e,Authors17h} to show the qualitative performance of QuDeC.

	\begin{figure*}[htp!]
		\centering
		\begin{minipage}{.16\textwidth}
			\centering
			\caption*{\emph{PSNR: 13.49\\SSIM: 0.55}}
			\vskip-5pt
			\includegraphics[width=2.7cm,height=1.58cm]{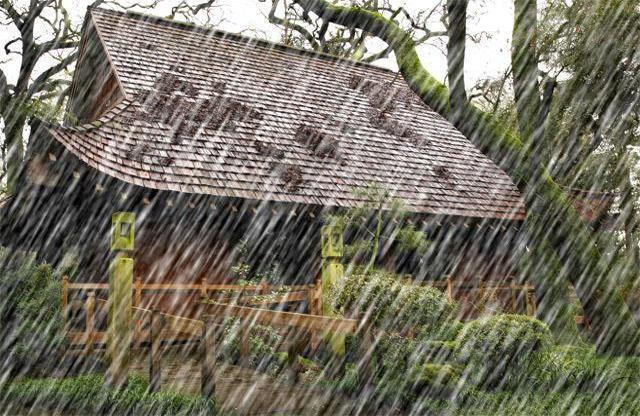}
			\captionsetup{labelformat=empty}
			\captionsetup{justification=centering}
		\end{minipage}  
		\begin{minipage}{.16\textwidth}
			\centering
			\caption*{\emph{PSNR:21.52 \\SSIM: 0.80}}
			\vskip-5pt
			\includegraphics[width=2.7cm,height=1.58cm]{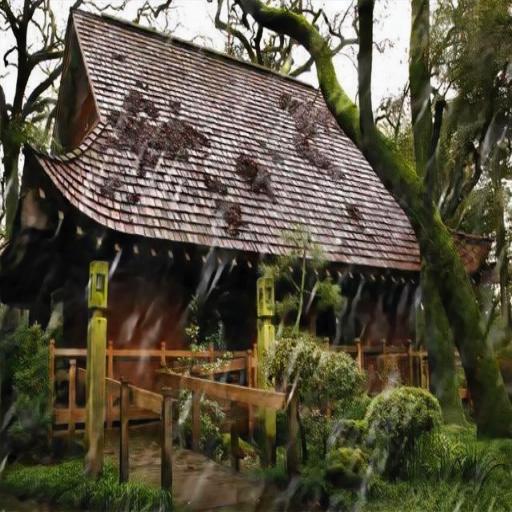}
			\captionsetup{labelformat=empty}
			\captionsetup{justification=centering}
		\end{minipage}
		\begin{minipage}{.16\textwidth}
			\centering
			\caption*{\emph{PSNR: 21.80 \\SSIM:0.78}}
			\vskip-5pt
			\includegraphics[width=2.7cm,height=1.58cm]{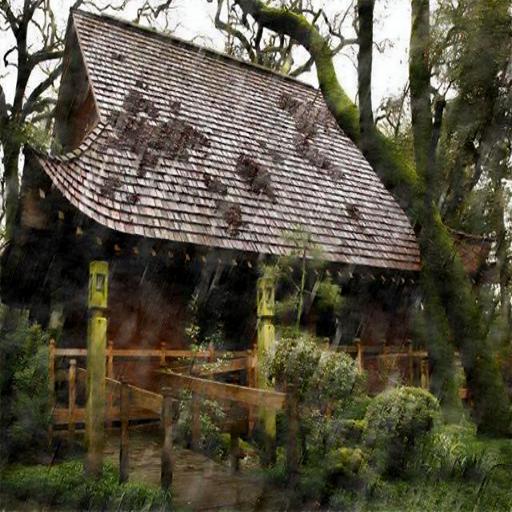}
			\captionsetup{labelformat=empty}
			\captionsetup{justification=centering}
		\end{minipage}
		\begin{minipage}{.16\textwidth}
			\centering
			\caption*{\emph{PSNR: 23.23 \\SSIM: 0.81}}
			\vskip-5pt
			\includegraphics[width=2.7cm,height=1.58cm]{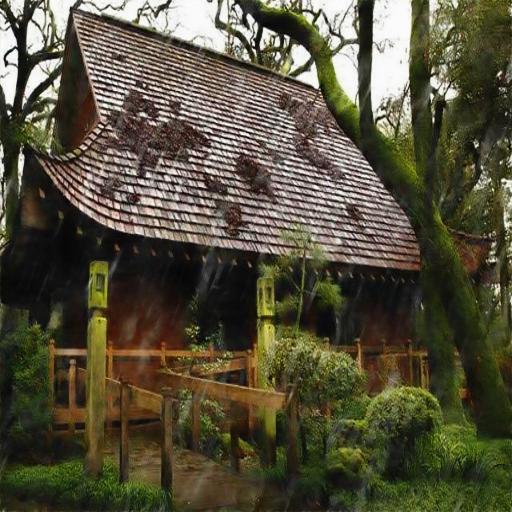}
			\captionsetup{labelformat=empty}
			\captionsetup{justification=centering}
		\end{minipage}
		\begin{minipage}{.16\textwidth}
			\centering
			\caption*{\emph{PSNR: \textbf{27.90} \\SSIM: \textbf{0.89}}}
			\vskip-5pt
			\includegraphics[width=2.7cm,height=1.58cm]{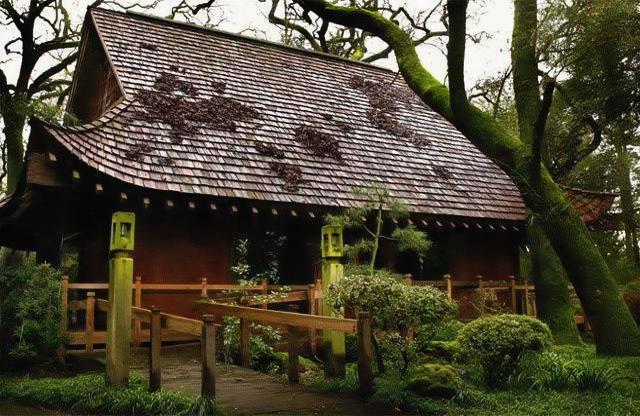}
			\captionsetup{labelformat=empty}
			\captionsetup{justification=centering}
		\end{minipage} 
		\begin{minipage}{.16\textwidth}
			\centering
			\caption*{\emph{PSNR: Inf\\SSIM: 1}}
			\vskip-5pt
			\includegraphics[width=2.7cm,height=1.58cm]{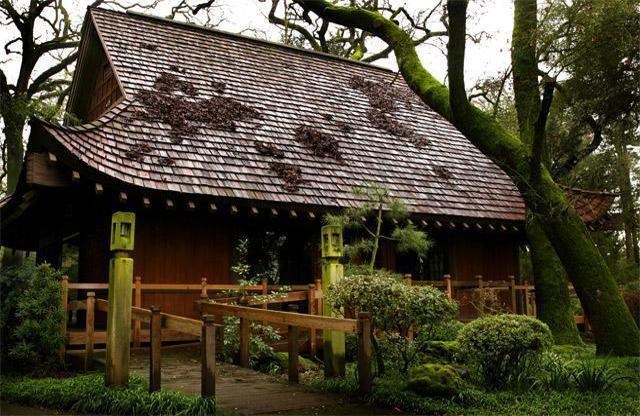}
			\captionsetup{labelformat=empty}
			\captionsetup{justification=centering}
		\end{minipage}\\ \vskip5pt
		\begin{minipage}{.16\textwidth}
			\centering
			\caption*{\emph{PSNR: 15.73\\SSIM: 0.70}}
			\vskip-5pt
			\includegraphics[width=2.7cm,height=1.58cm]{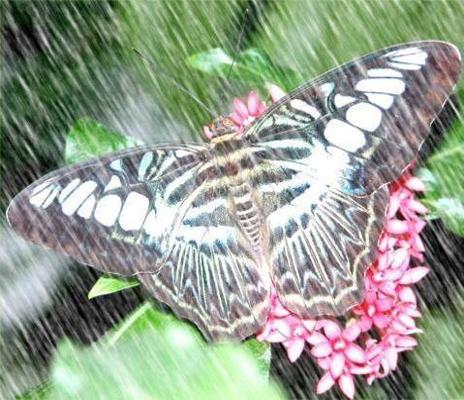}
			\captionsetup{labelformat=empty}
			\captionsetup{justification=centering}
		\end{minipage}  
		\begin{minipage}{.16\textwidth}
			\centering
			\caption*{\emph{PSNR:22.01 \\SSIM: 0.81}}
			\vskip-5pt
			\includegraphics[width=2.7cm,height=1.58cm]{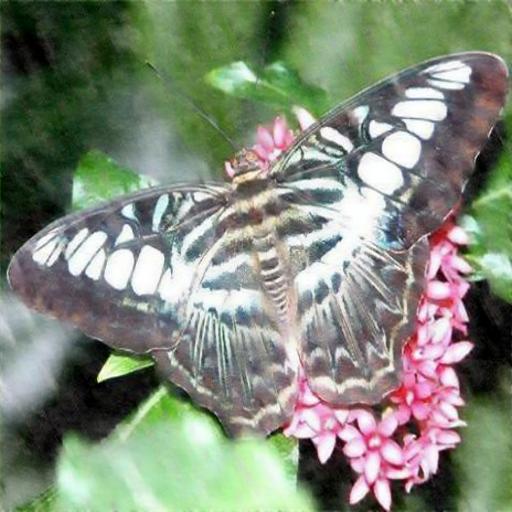}
			\captionsetup{labelformat=empty}
			\captionsetup{justification=centering}
		\end{minipage}
		\begin{minipage}{.16\textwidth}
			\centering
			\caption*{\emph{PSNR: 24.14 \\SSIM: 0.86}}
			\vskip-5pt
			\includegraphics[width=2.7cm,height=1.58cm]{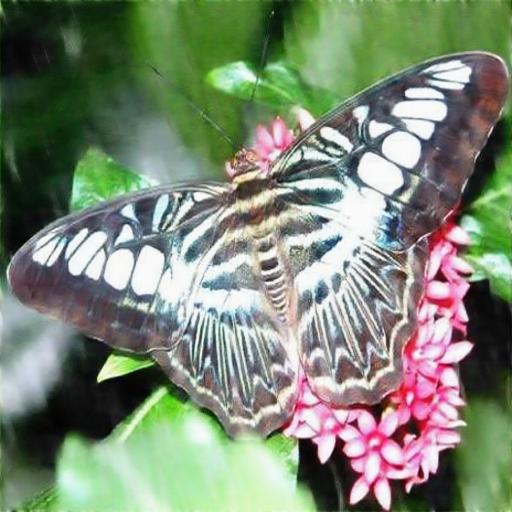}
			\captionsetup{labelformat=empty}
			\captionsetup{justification=centering}
		\end{minipage}
		\begin{minipage}{.16\textwidth}
			\centering
			\caption*{\emph{PSNR: 22.71 \\SSIM:0.85}}
			\vskip-5pt
			\includegraphics[width=2.7cm,height=1.58cm]{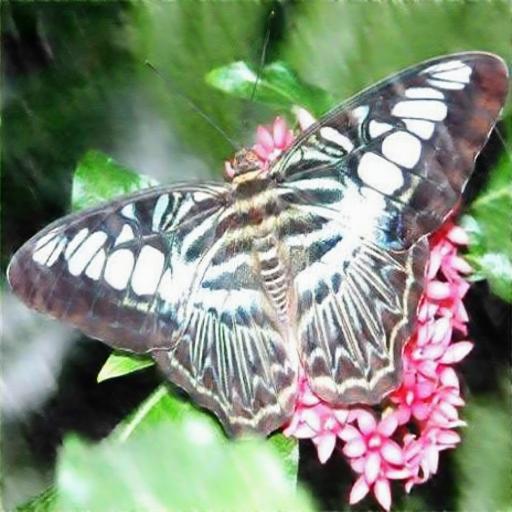}
			\captionsetup{labelformat=empty}
			\captionsetup{justification=centering}
		\end{minipage}
		\begin{minipage}{.16\textwidth}
			\centering
			\caption*{\emph{PSNR: \textbf{28.89} \\SSIM: \textbf{0.92}}}
			\vskip-5pt
			\includegraphics[width=2.7cm,height=1.58cm]{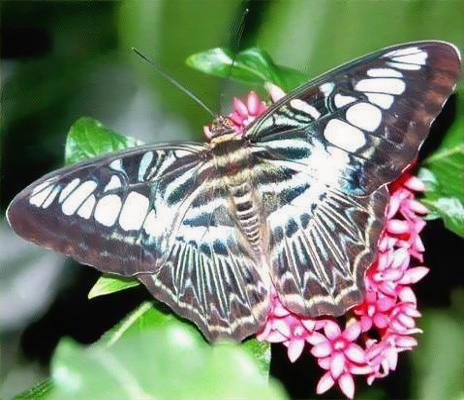}
			\captionsetup{labelformat=empty}
			\captionsetup{justification=centering}
		\end{minipage} 
		\begin{minipage}{.16\textwidth}
			\centering
			\caption*{\emph{PSNR: Inf\\SSIM: 1}}
			\vskip-5pt
			\includegraphics[width=2.7cm,height=1.58cm]{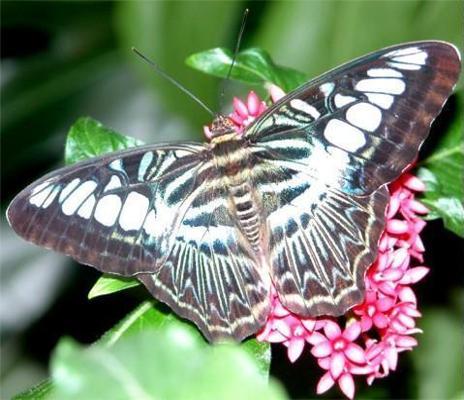}
			\captionsetup{labelformat=empty}
			\captionsetup{justification=centering}
		\end{minipage}\\ \vskip5pt
		\begin{minipage}{.16\textwidth}
			\centering
			\caption*{\emph{PSNR: 14.76\\SSIM: 0.72}}
			\vskip-5pt
			\includegraphics[width=2.7cm,height=1.58cm]{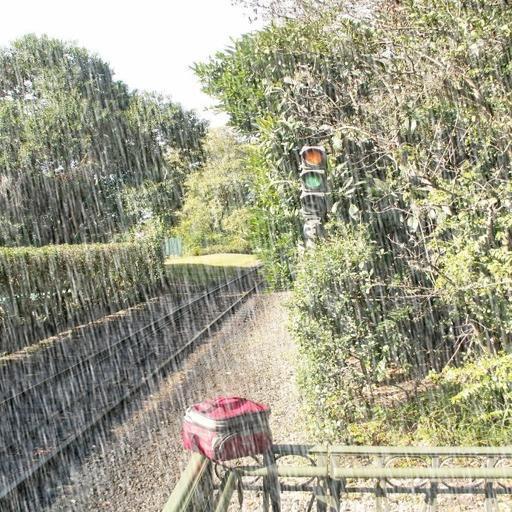}
			\captionsetup{labelformat=empty}
			\captionsetup{justification=centering}
		\end{minipage}  
		\begin{minipage}{.16\textwidth}
			\centering
			\caption*{\emph{PSNR:19.10 \\SSIM: 0.80}}
			\vskip-5pt
			\includegraphics[width=2.7cm,height=1.58cm]{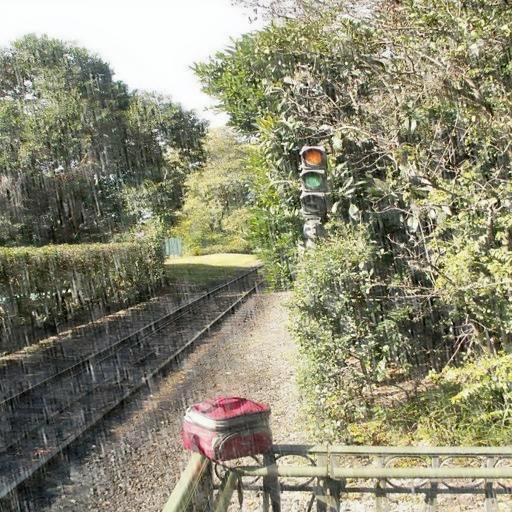}
			\captionsetup{labelformat=empty}
			\captionsetup{justification=centering}
		\end{minipage}
		\begin{minipage}{.16\textwidth}
			\centering
			\caption*{\emph{PSNR: 21.37 \\SSIM: 0.82}}
			\vskip-5pt
			\includegraphics[width=2.7cm,height=1.58cm]{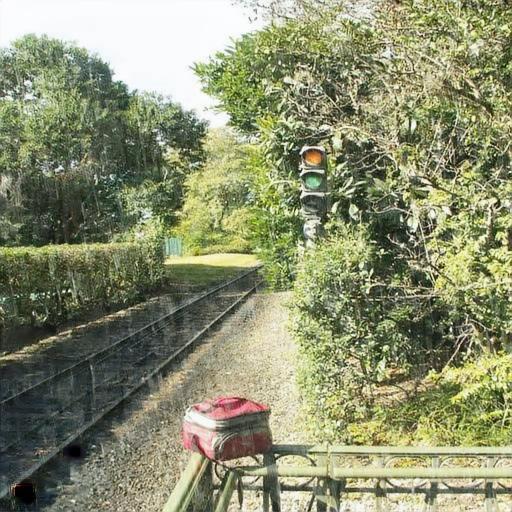}
			\captionsetup{labelformat=empty}
			\captionsetup{justification=centering}
		\end{minipage}
		\begin{minipage}{.16\textwidth}
			\centering
			\caption*{\emph{PSNR: 21.95 \\SSIM:0.86}}
			\vskip-5pt
			\includegraphics[width=2.7cm,height=1.58cm]{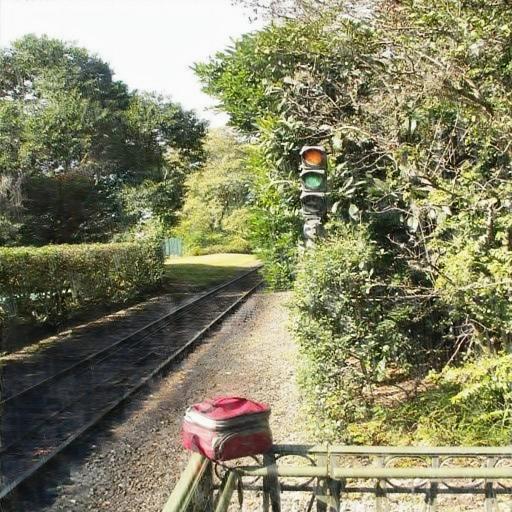}
			\captionsetup{labelformat=empty}
			\captionsetup{justification=centering}
		\end{minipage}
		\begin{minipage}{.16\textwidth}
			\centering
			\caption*{\emph{PSNR: \textbf{24.01} \\SSIM: \textbf{0.89}}}
			\vskip-5pt
			\includegraphics[width=2.7cm,height=1.58cm]{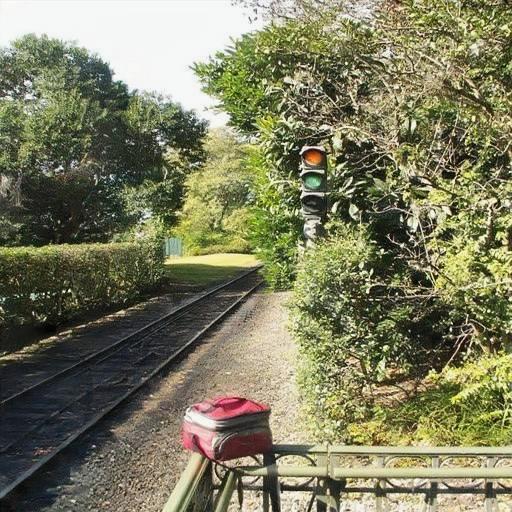}
			\captionsetup{labelformat=empty}
			\captionsetup{justification=centering}
		\end{minipage} 
		\begin{minipage}{.16\textwidth}
			\centering
			\caption*{\emph{PSNR: Inf\\SSIM: 1}}
			\vskip-5pt
			\includegraphics[width=2.7cm,height=1.58cm]{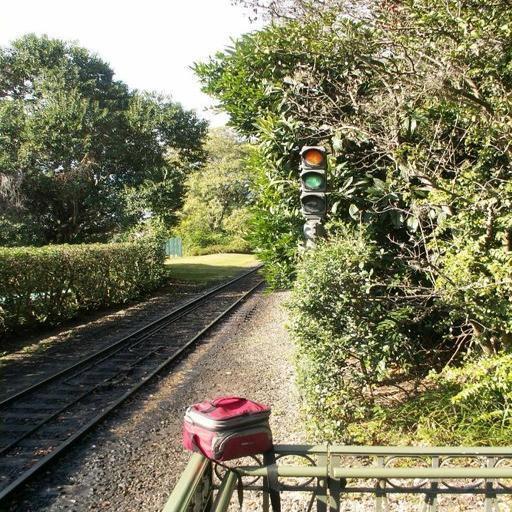}
			\captionsetup{labelformat=empty}
			\captionsetup{justification=centering}
		\end{minipage}\\ \vskip5pt
		\begin{minipage}{.16\textwidth}
			\centering
			\caption*{\emph{PSNR: 14.75\\SSIM: 0.59}}
			\vskip-5pt
			\includegraphics[width=2.7cm,height=1.58cm]{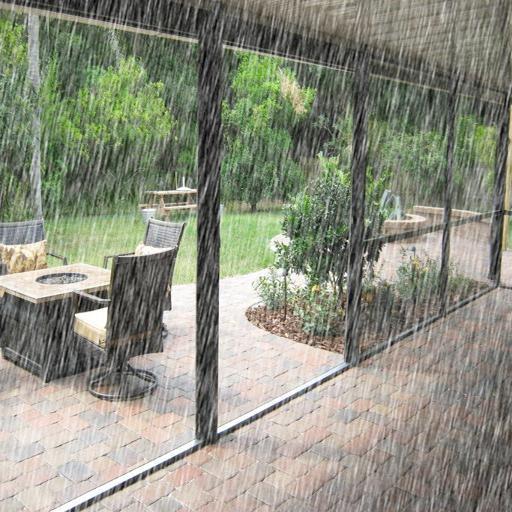}
			\captionsetup{labelformat=empty}
			\captionsetup{justification=centering}
		\end{minipage}  
		\begin{minipage}{.16\textwidth}
			\centering
			\caption*{\emph{PSNR:21.22 \\SSIM: 0.76}}
			\vskip-5pt
			\includegraphics[width=2.7cm,height=1.58cm]{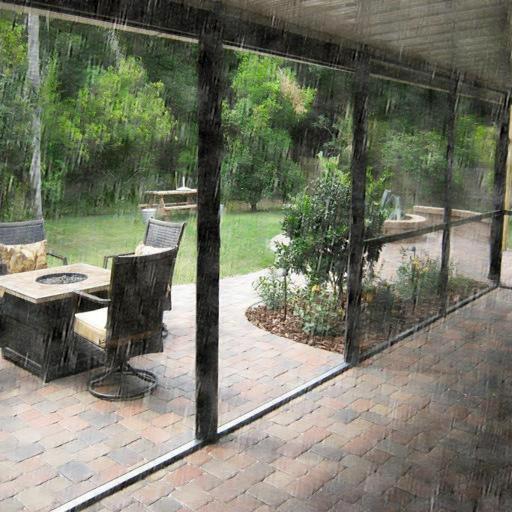}
			\captionsetup{labelformat=empty}
			\captionsetup{justification=centering}
		\end{minipage}
		\begin{minipage}{.16\textwidth}
			\centering
			\caption*{\emph{PSNR: 25.78 \\SSIM: 0.83}}
			\vskip-5pt
			\includegraphics[width=2.7cm,height=1.58cm]{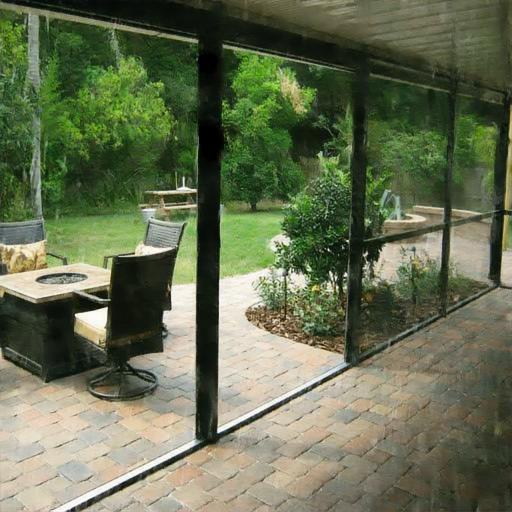}
			\captionsetup{labelformat=empty}
			\captionsetup{justification=centering}
		\end{minipage}
		\begin{minipage}{.16\textwidth}
			\centering
			\caption*{\emph{PSNR: 25.54 \\SSIM:0.85}}
			\vskip-5pt
			\includegraphics[width=2.7cm,height=1.58cm]{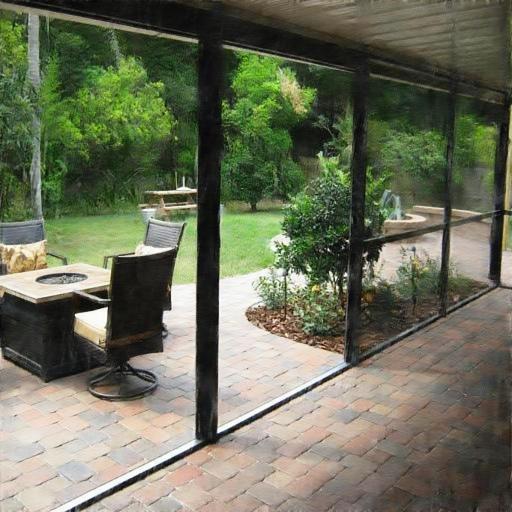}
			\captionsetup{labelformat=empty}
			\captionsetup{justification=centering}
		\end{minipage}
		\begin{minipage}{.16\textwidth}
			\centering
			\caption*{\emph{PSNR: \textbf{27.29} \\SSIM: \textbf{0.88}}}
			\vskip-5pt
			\includegraphics[width=2.7cm,height=1.58cm]{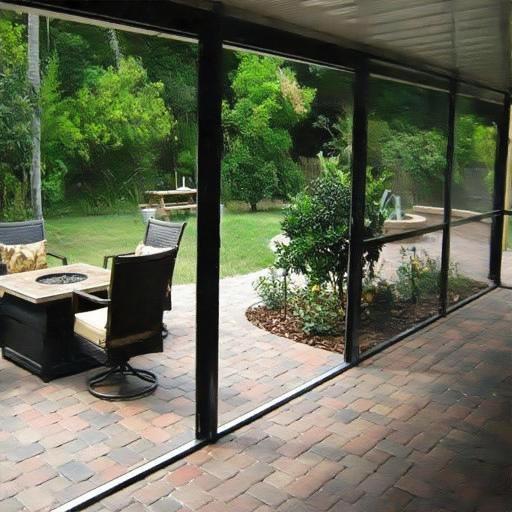}
			\captionsetup{labelformat=empty}
			\captionsetup{justification=centering}
		\end{minipage} 
		\begin{minipage}{.16\textwidth}
			\centering
			\caption*{\emph{PSNR: Inf\\SSIM: 1}}
			\vskip-5pt
			\includegraphics[width=2.7cm,height=1.58cm]{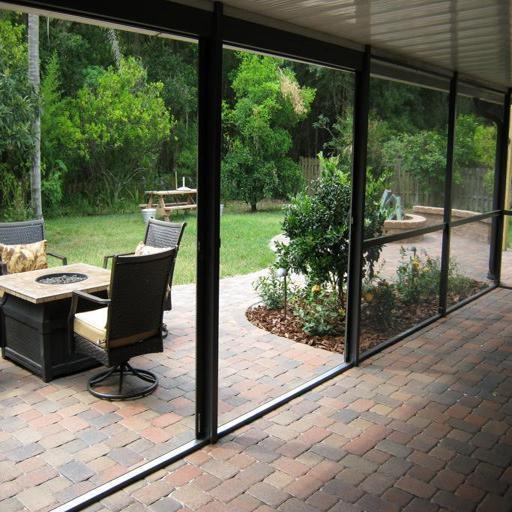}
			\captionsetup{labelformat=empty}
			\captionsetup{justification=centering}
		\end{minipage}\\ \vskip5pt
		\begin{minipage}{.16\textwidth}
			\centering
			\caption*{\emph{PSNR: 21.50\\SSIM: 0.84}}
			\vskip-5pt
			\includegraphics[width=2.7cm,height=1.58cm]{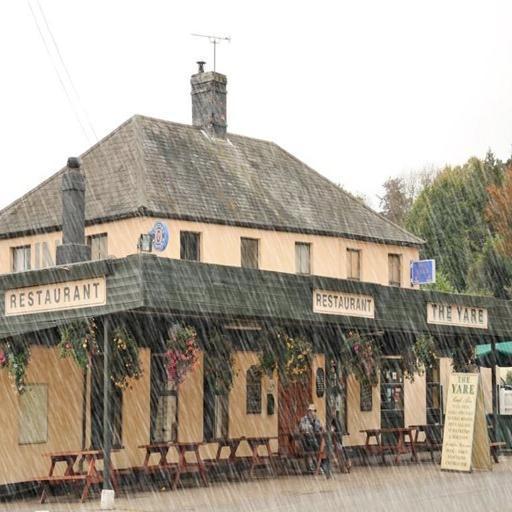}
			\captionsetup{labelformat=empty}
			\captionsetup{justification=centering}
		\end{minipage}  
		\begin{minipage}{.16\textwidth}
			\centering
			\caption*{\emph{PSNR:28.77 \\SSIM: 0.92}}
			\vskip-5pt
			\includegraphics[width=2.7cm,height=1.58cm]{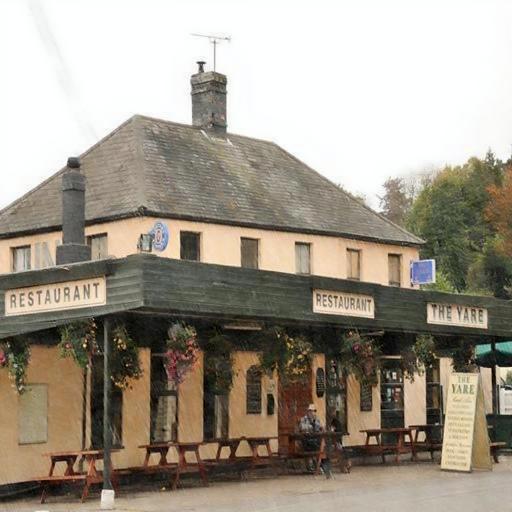}
			\captionsetup{labelformat=empty}
			\captionsetup{justification=centering}
		\end{minipage}
		\begin{minipage}{.16\textwidth}
			\centering
			\caption*{\emph{PSNR: 29.83 \\SSIM: 0.94}}
			\vskip-5pt
			\includegraphics[width=2.7cm,height=1.58cm]{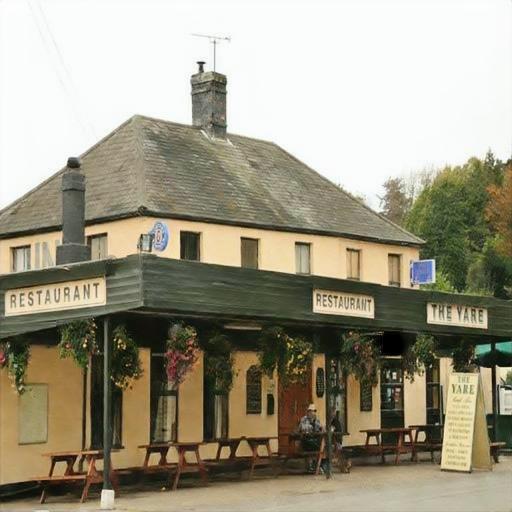}
			\captionsetup{labelformat=empty}
			\captionsetup{justification=centering}
		\end{minipage}
		\begin{minipage}{.16\textwidth}
			\centering
			\caption*{\emph{PSNR: 28.63 \\SSIM:0.93}}
			\vskip-5pt
			\includegraphics[width=2.7cm,height=1.58cm]{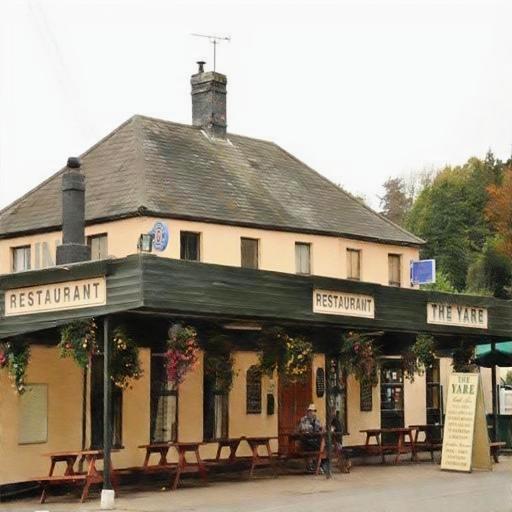}
			\captionsetup{labelformat=empty}
			\captionsetup{justification=centering}
		\end{minipage}
		\begin{minipage}{.16\textwidth}
			\centering
			\caption*{\emph{PSNR: \textbf{29.27} \\SSIM: \textbf{0.96}}}
			\vskip-5pt
			\includegraphics[width=2.7cm,height=1.58cm]{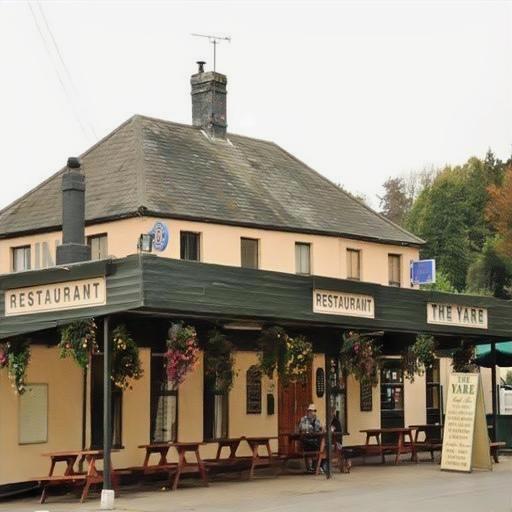}
			\captionsetup{labelformat=empty}
			\captionsetup{justification=centering}
		\end{minipage} 
		\begin{minipage}{.16\textwidth}
			\centering
			\caption*{\emph{PSNR: Inf\\SSIM: 1}}
			\vskip-5pt
			\includegraphics[width=2.7cm,height=1.58cm]{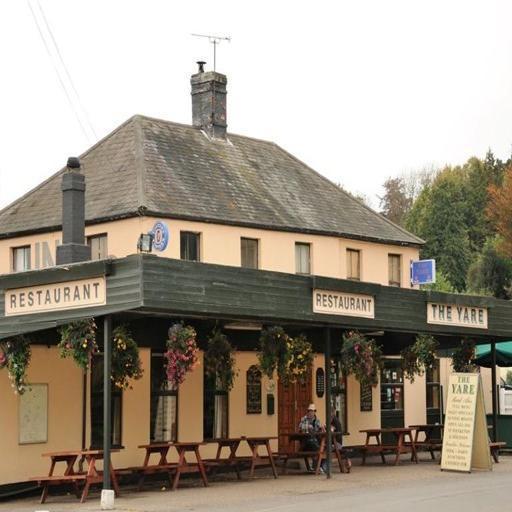}
			\captionsetup{labelformat=empty}
			\captionsetup{justification=centering}
		\end{minipage}\\ \vskip5pt
		\begin{minipage}{.16\textwidth}
			\centering
			\caption*{\emph{PSNR:20.57 \\SSIM: 0.83}}
			\vskip-5pt
			\includegraphics[width=2.7cm,height=1.58cm]{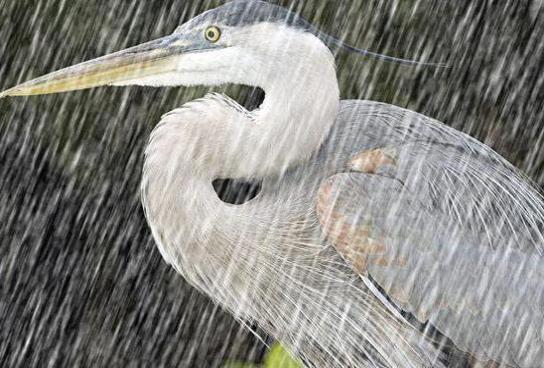}
			\captionsetup{labelformat=empty}
			\captionsetup{justification=centering}
			\caption*{\emph{Rainy Image} \\ \quad\\ \quad}
		\end{minipage} 
		\begin{minipage}{.16\textwidth}
			\centering
			\caption*{\emph{PSNR:23.48 \\SSIM: 0.73}}
			\vskip-5pt
			\includegraphics[width=2.7cm,height=1.58cm]{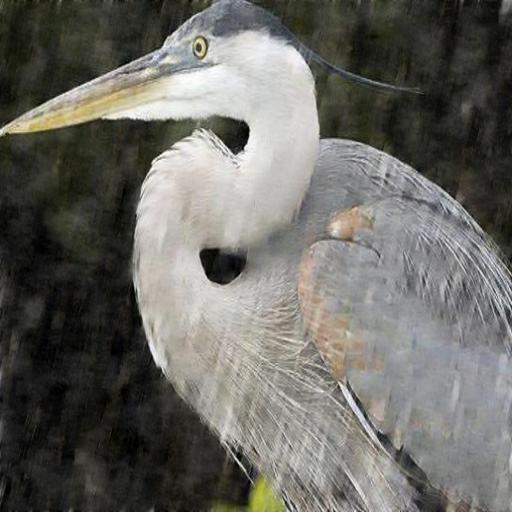}
			\captionsetup{labelformat=empty}
			\captionsetup{justification=centering}
			\caption*{\emph{DDN \\ \cite{Authors17f}(CVPR'17)} \\ \quad}
		\end{minipage} 
		\begin{minipage}{.16\textwidth}
			\centering
			\caption*{\emph{PSNR:26.05\\SSIM:0.84}}
			\vskip-5pt
			\includegraphics[width=2.7cm,height=1.58cm]{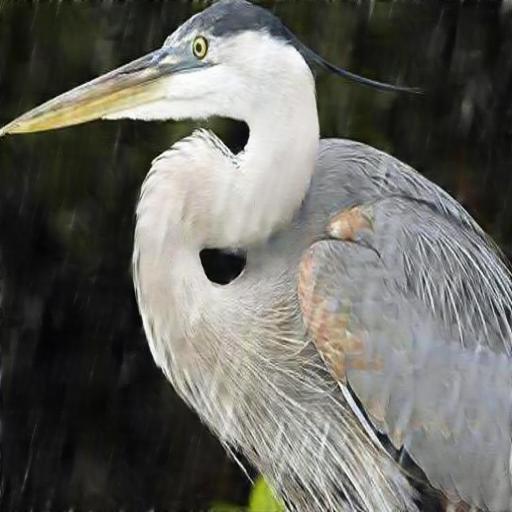}
			\captionsetup{labelformat=empty}
			\captionsetup{justification=centering}
			\caption*{\emph{RESCAN \cite{Authors18d}(ECCV'18)} \\ \quad}
		\end{minipage} 
		\begin{minipage}{.16\textwidth}
			\centering
			\caption*{\emph{PSNR: 29.21 \\SSIM:0.88}}
			\vskip-5pt
			\includegraphics[width=2.7cm,height=1.58cm]{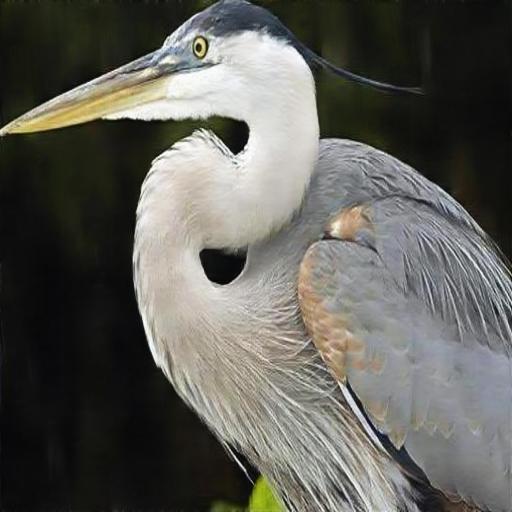}
			\captionsetup{labelformat=empty}
			\captionsetup{justification=centering}
			\caption*{\emph{DID-MDN \cite{Authors18}(CVPR'18)} \\ \quad}
		\end{minipage} 
		\begin{minipage}{.16\textwidth}
			\centering
			\caption*{\emph{PSNR: \textbf{29.94}\\SSIM:\textbf{0.89}}}
			\vskip-5pt
			\includegraphics[width=2.7cm,height=1.58cm]{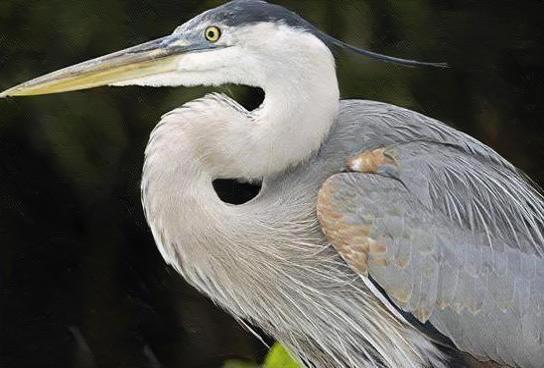}
			\captionsetup{labelformat=empty}
			\captionsetup{justification=centering}
			\caption*{\emph{QuDeC} \\ ours \quad\\ \quad}
		\end{minipage}
		\begin{minipage}{.16\textwidth}
			\centering
			\caption*{\emph{PSNR: Inf \\SSIM:1}}
			\vskip-5pt
			\includegraphics[width=2.7cm,height=1.58cm]{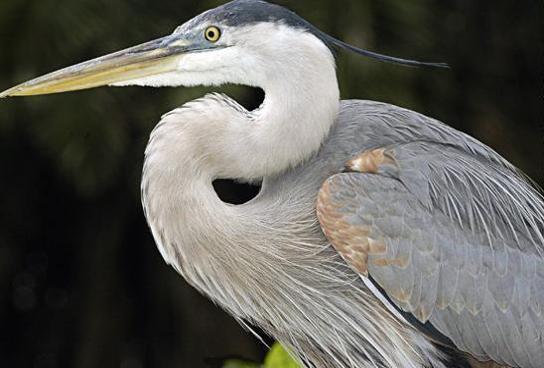}
			\captionsetup{labelformat=empty}
			\captionsetup{justification=centering}
			\caption*{\emph{Ground Truth} \\ \quad\\ \quad}
		\end{minipage} \\
		\vskip-20pt
		\caption{Rain-streak removal results on sample images from the synthetic images consisting of different rain levels (low, medium and heavy) and directions.}\label{Fig:exp4}
	\end{figure*}
	
	\begin{table*}[htp!]
		\begin{center}
			\caption{\large{Quantitative results evaluated in terms of average SSIM and PSNR (dB) (PSNR$|$SSIM).}}
			\resizebox{\textwidth}{!}{
				\label{syn_comp}
				\begin{tabular}{|c|c|c|c|c|c|c|c|c|}
					\hline
					Dataset & Testset & \begin{tabular}[c]{@{}c@{}}Fu et al.\\ \cite{Authors17d}(TIP'17)\end{tabular} & \begin{tabular}[c]{@{}c@{}}DDN\\ \cite{Authors17f}(CVPR'17)\end{tabular} & \begin{tabular}[c]{@{}c@{}}JORDER\\ \cite{Authors17h}(CVPR'17)\end{tabular} & \begin{tabular}[c]{@{}c@{}}RESCAN\\ \cite{Authors18d}(CVPR'18)\end{tabular}                           & \begin{tabular}[c]{@{}c@{}}DID-MDN\\ \cite{Authors18}(CVPR'18)\end{tabular} & \begin{tabular}[c]{@{}c@{}c@{}}UMRL+\\ cycle spinning\\\cite{Authors19} (CVPR'19)\end{tabular} & \begin{tabular}[c]{@{}c@{}}QuDeC\\ ours\end{tabular} \\ \hline
					& \textit{Test-1}   & 22.07$|$0.84 & 27.33$|$0.90 & 24.32$|$0.86 & 27.19$|$0.87 & {\color[HTML]{009901} 27.95$|$0.91} & {\color[HTML]{00009B} 29.77$|$0.92} & {\color[HTML]{CB0000} 30.43$|$0.93}                    \\ \cline{2-9} 
					\multirow{-2}{*}{DIDMDN\cite{Authors18}} & \textit{Test-2}   & 19.73$|$0.83 & 25.63$|$0.88 & 22.26$|$0.84 & 25.65$|$0.88 & {\color[HTML]{009901} 26.08$|$0.90} & {\color[HTML]{00009B} 26.67$|$0.92} & {\color[HTML]{CB0000} 26.72$|$0.92}                    \\ \hline
					Rain800~\cite{Authors17e} & \textit{Rain800}  & 18.95$|$0.78 & 21.33$|$0.80 & 21.13$|$0.81 & {\color[HTML]{009901} 24.37$|$0.84} & 23.57$|$0.87 & {\color[HTML]{00009B} 24.52$|$0.86} & {\color[HTML]{CB0000} 24.61$|$0.86} \\ \hline
					JORDER~\cite{Authors17h} & \textit{Rain200H} & 21.83$|$0.81  & 23.24$|$0.83 & 24.02$|$0.86 & {\color[HTML]{009901} 25.97$|$0.90} & 23.43$|$0.86  & {\color[HTML]{00009B} 26.38$|$0.92} & {\color[HTML]{CB0000} 26.74$|$0.93} \\ \hline
				\end{tabular}
			}
			\\
			\vskip3pt
			Values highlighted in ${\color[HTML]{CB0000}red\:color}$- indicate the best performance, ${\color[HTML]{00009B}blue \:color}-$ indicate the second best performance, ${\color[HTML]{009901}green\:color}-$ indicate the third best performance among the de-raining methods on the test datasets.
		\end{center}
	\end{table*}
	
	\subsection{Training Details}
	The $(y,x,s)$ rainy-clean-distortion label image pairs  are used to train QuDeC using the loss $\mathcal{L}$. The Adam optimizer with the batch size of 1 is used to train the network. Learning rate is set equal to 0.0002 for the first 20 epochs and 0.0001 for the remaining epochs. During training initially $\lambda_1$, $\lambda_{cs}$ and $\lambda_2$ are set equal to 0.1, 0.1 and 1.0, respectively, but when the mean of all values in the confidences maps $c_{\x1},\;c_{\x2}$ and $c_{\x4}$ is greater than 0.8 then $\lambda_1$ is set equal to 0.03. QuDeC is trained for 60 epochs. During inference given a rainy image $y$, QuDeC estimates the residual map and its corresponding confidence map at three different scales, $\:\{\hat{x},\hat{r}_{\x1},c_{\x1}\} $ (at the original input size), $\:\{\hat{x}_{\x2},\hat{r}_{\x2},c_{\x2}\} $, and $\:\{\hat{x}_{\x4},\hat{r}_{\x4},c_{\x4}\} $ (at 0.25 scale of the input size) along with the location-quality-label-map $\hat{s}$.

	\begin{figure*}[htp!]
		\begin{center}
			\includegraphics[width=0.195\textwidth,height=0.12\textwidth]{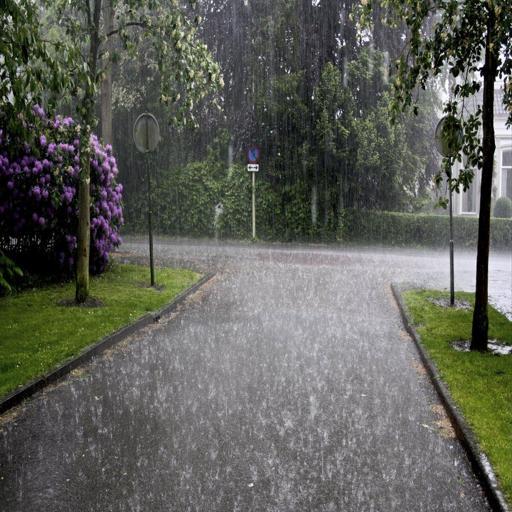}
			\includegraphics[width=0.195\textwidth,height=0.12\textwidth]{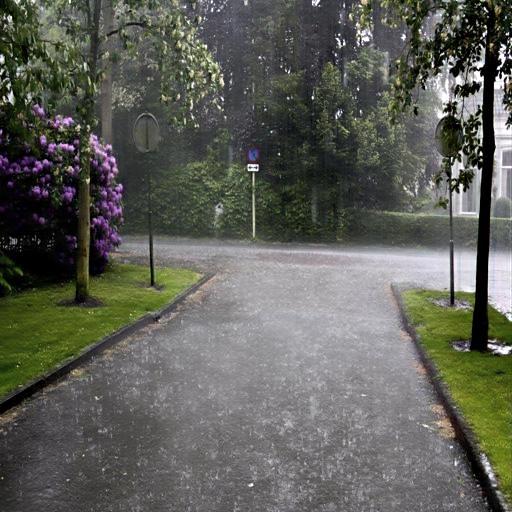}
			\includegraphics[width=0.195\textwidth,height=0.12\textwidth]{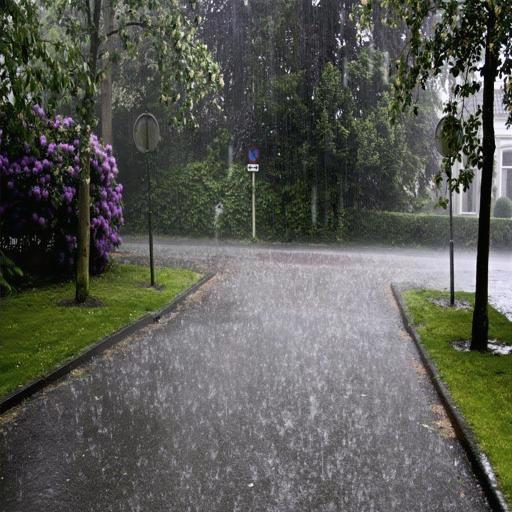}
			\includegraphics[width=0.195\textwidth,height=0.12\textwidth]{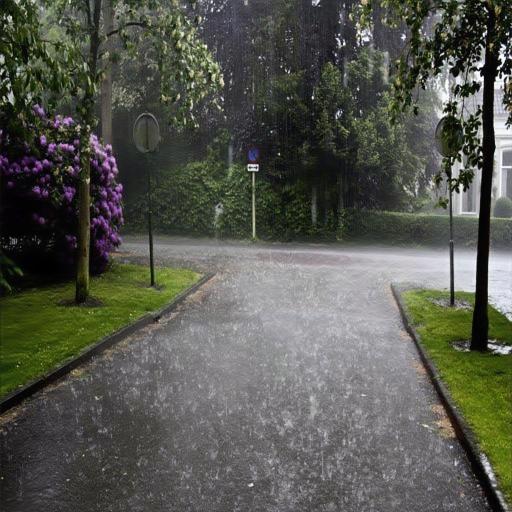}
			\includegraphics[width=0.195\textwidth,height=0.12\textwidth]{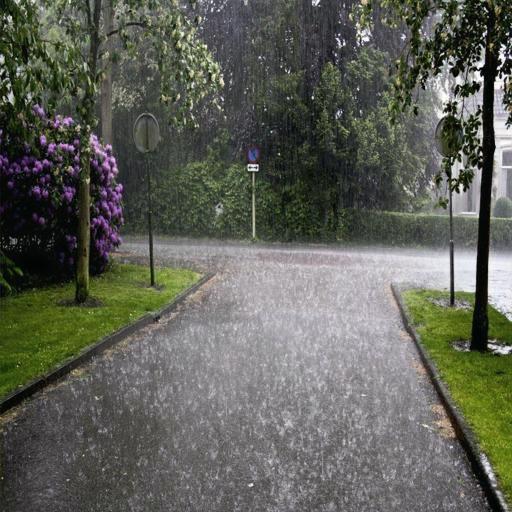}\\ \vskip3pt
			\includegraphics[width=0.195\textwidth,height=0.12\textwidth]{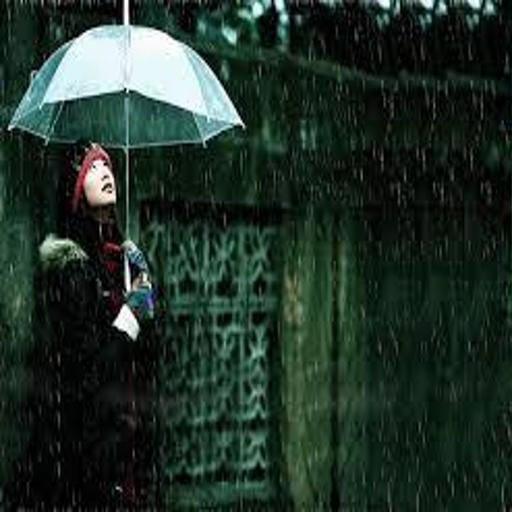}
			\includegraphics[width=0.195\textwidth,height=0.12\textwidth]{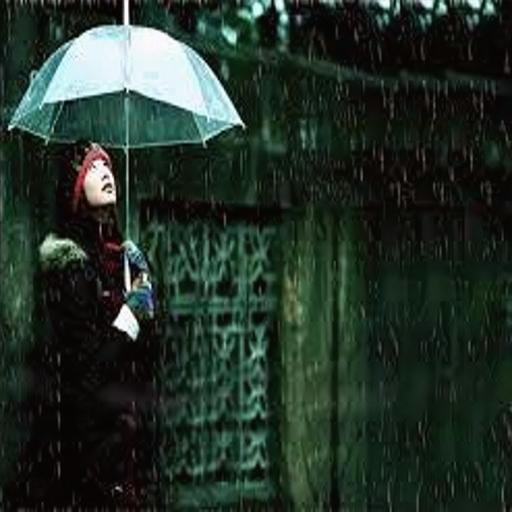}
			\includegraphics[width=0.195\textwidth,height=0.12\textwidth]{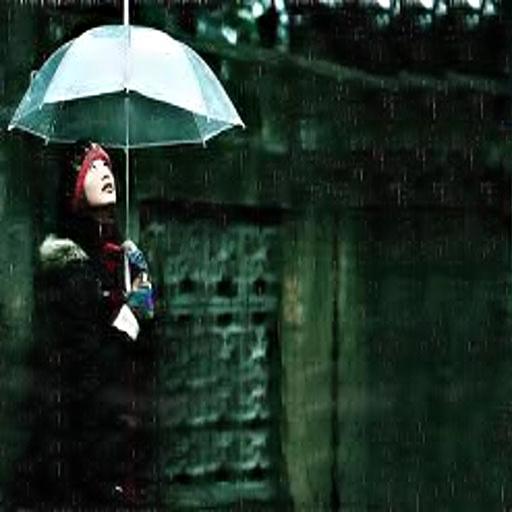}
			\includegraphics[width=0.195\textwidth,height=0.12\textwidth]{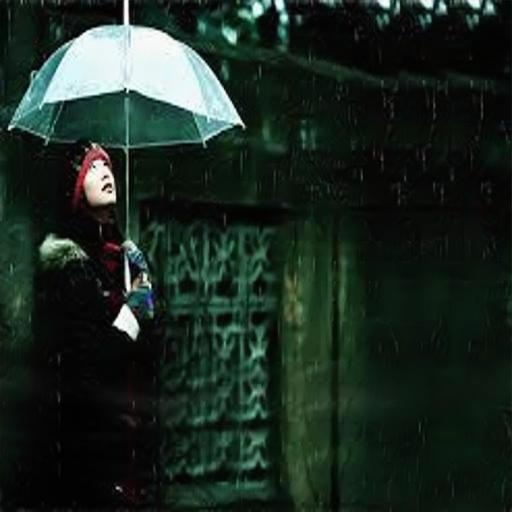}
			\includegraphics[width=0.195\textwidth,height=0.12\textwidth]{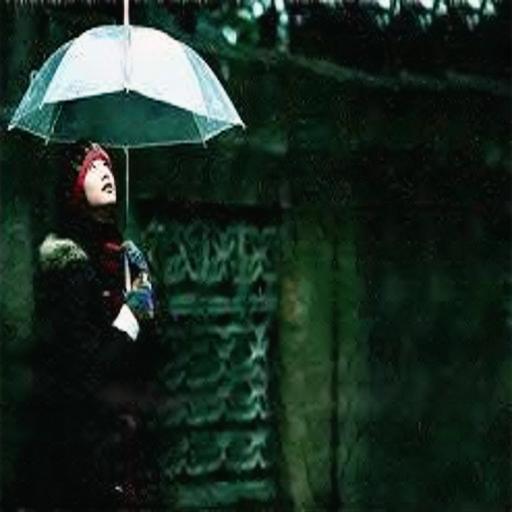}\\ \vskip3pt
			\includegraphics[width=0.195\textwidth,height=0.12\textwidth]{55_rainy.jpg}
			\includegraphics[width=0.195\textwidth,height=0.12\textwidth]{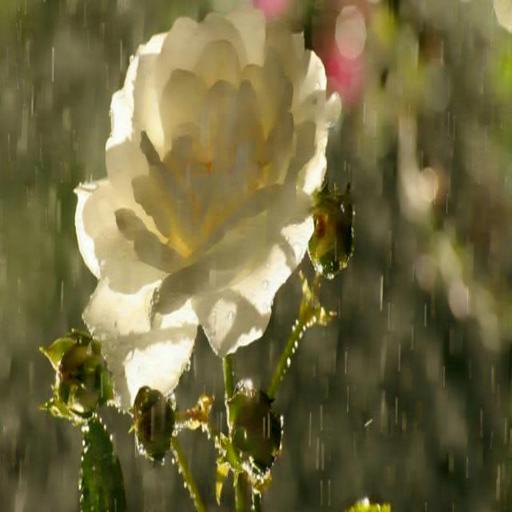}
			\includegraphics[width=0.195\textwidth,height=0.12\textwidth]{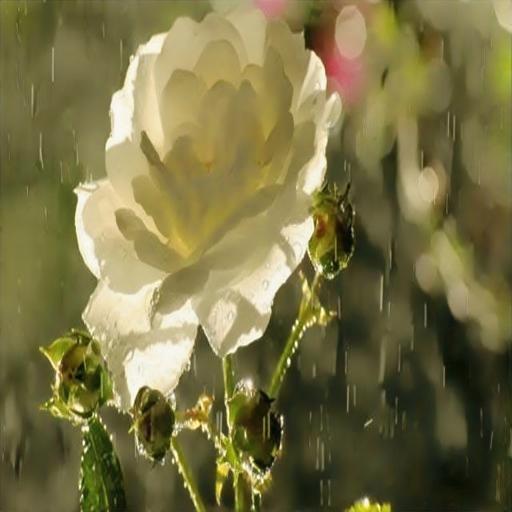}
			\includegraphics[width=0.195\textwidth,height=0.12\textwidth]{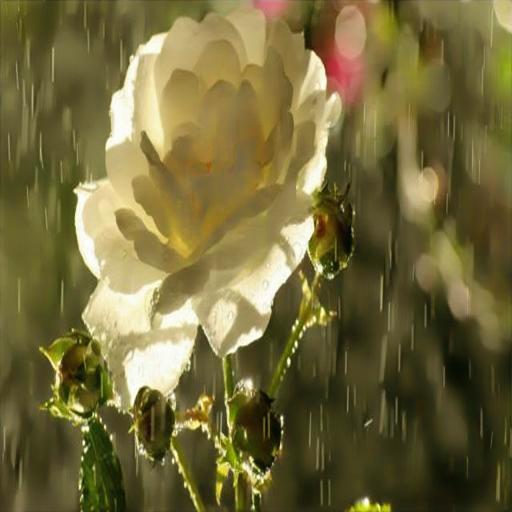}
			\includegraphics[width=0.195\textwidth,height=0.12\textwidth]{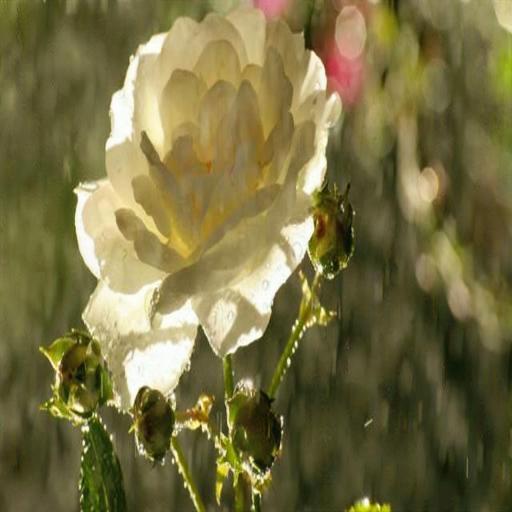}\\ \vskip3pt
			\includegraphics[width=0.195\textwidth,height=0.12\textwidth]{63_rainy.jpg}
			\includegraphics[width=0.195\textwidth,height=0.12\textwidth]{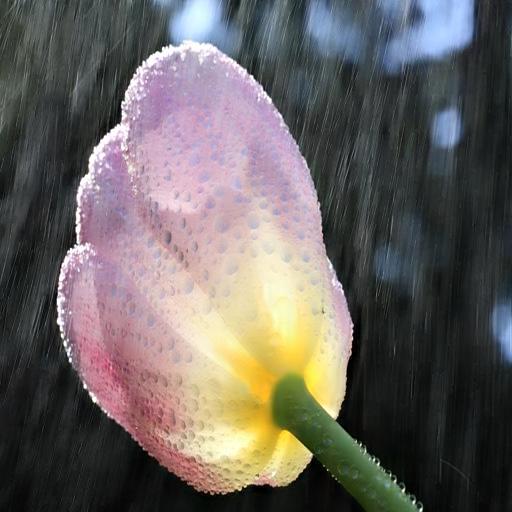}
			\includegraphics[width=0.195\textwidth,height=0.12\textwidth]{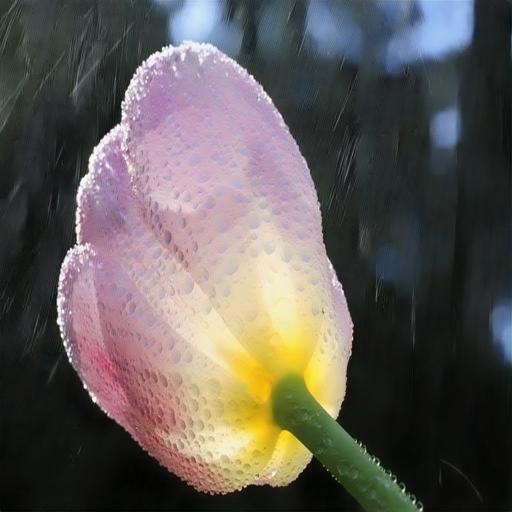}
			\includegraphics[width=0.195\textwidth,height=0.12\textwidth]{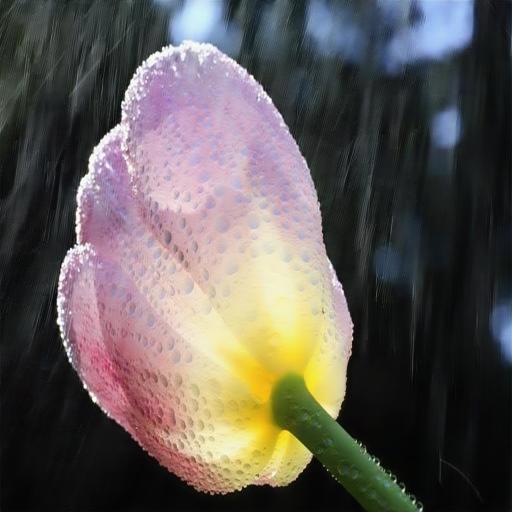}
			\includegraphics[width=0.195\textwidth,height=0.12\textwidth]{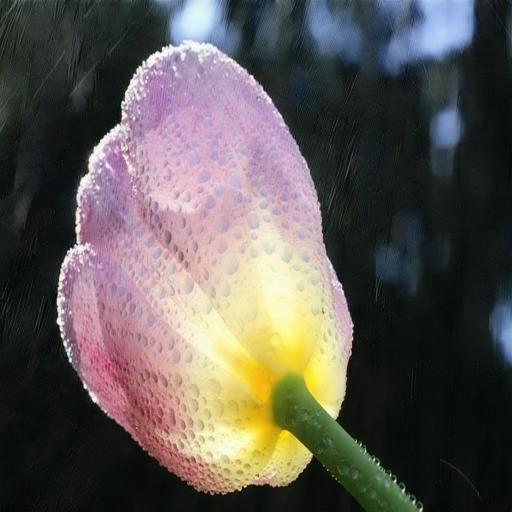}\\ \vskip3pt
			\includegraphics[width=0.195\textwidth,height=0.12\textwidth]{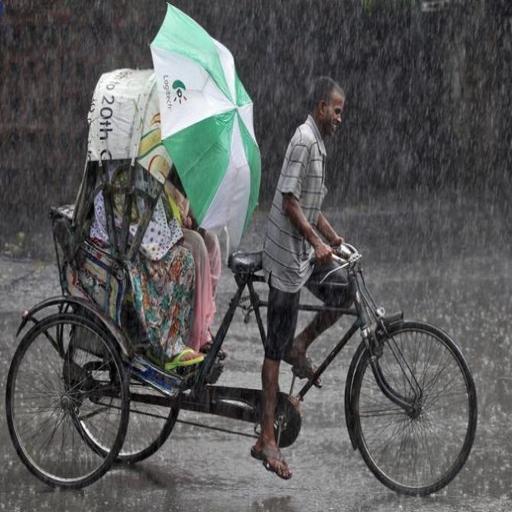}
			\includegraphics[width=0.195\textwidth,height=0.12\textwidth]{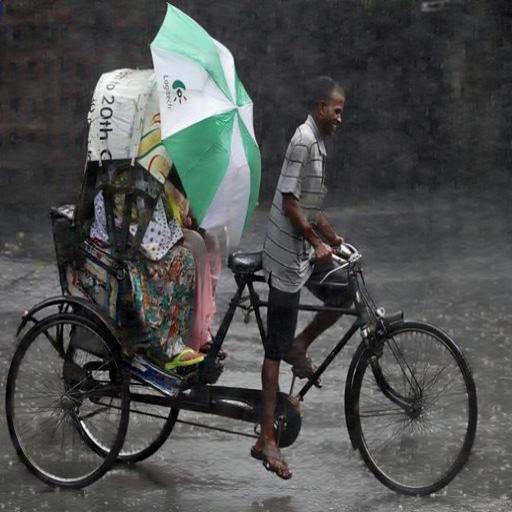}
			\includegraphics[width=0.195\textwidth,height=0.12\textwidth]{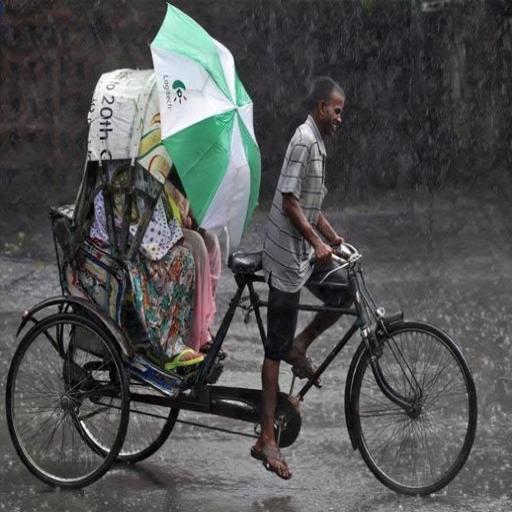}
			\includegraphics[width=0.195\textwidth,height=0.12\textwidth]{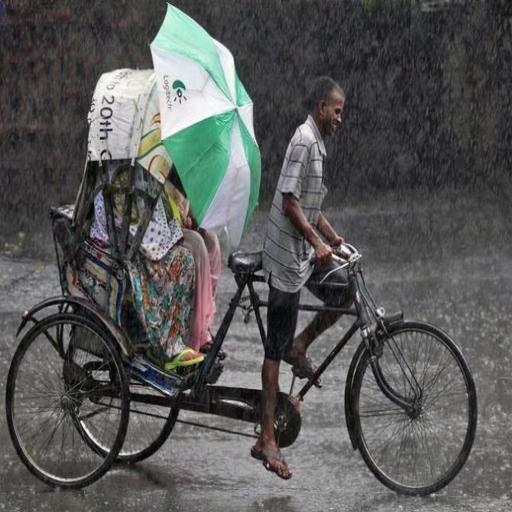}
			\includegraphics[width=0.195\textwidth,height=0.12\textwidth]{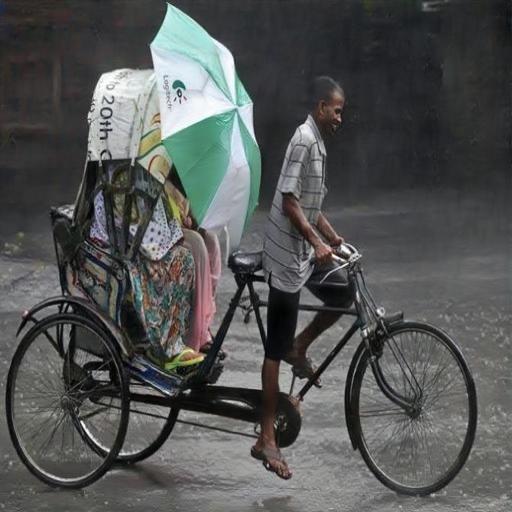}\\ 
			\textit{ Rainy  \hskip70pt DDN \hskip70pt RESCAN \hskip60pt DID-MDN  \hskip60pt QuDeC\\}
			\textit{Image \hskip60pt \cite{Authors17f}(CVPR'17) \hskip40pt \cite{Authors18d}(ECCV'18) \hskip40pt \cite{Authors18}(CVPR'18) \hskip50pt (Ours)\\ }
			\caption{De-raining results on sample real-world images.}
			\label{Fig:exp6}
		\end{center}
	\end{figure*}
	
	\section{Experimental Results}\label{sec:results}
	In this section, we evaluate the performance of our method on both synthetic and real images.  Peak-Signal-to-Noise Ratio (PSNR) and Structural Similarity index (SSIM) \cite{SSIM} measures are used to compare the performance of different methods on synthetic images.   We visually inspect the performance of different methods on real images, as we don't have the ground truth clean images. The performance of the proposed QuDeC method is compared against the following recent state-of-the-art methods:\\ 
	(a) Fu et al.\cite{Authors17d} CNN method (TIP'17),\\ 
	(b) Joint Rain Detection and Removal (JORDER) \cite{Authors17h} (CVPR’17),\\ 
	(c) Deep detailed Network (DDN)\cite{Authors17f} (CVPR'17),\\ 
	(d) Density-aware Image De-raining method using a Multistream
	Dense Network (DID-MDN) \cite{Authors18} (CVPR'18),\\ 
	(e) REcurrent SE Context Aggregation Net (RESCAN) \cite{Authors18d} (ECCV'18)\\
	(f) Uncertainty guided Multi-scale Residual Learning (UMRL) network \cite{Authors19} (CVPR'19).
	
	\subsection{Results on Synthetic Test Images}
	The proposed QuDeC method based on cycle spinning \cite{Authors19} is also compared against the state-of-the-art algorithms qualitatively and quantitatively. Table \ref{syn_comp} shows the quantitative performance of our method.  As it can be seen from this table, our method clearly out-performs the present state-of-the-art image de-raining algorithms. On average, QuDeC outperformes the methods like RESCAN~\cite{Authors18d} and DIDMDN~\cite{Authors18} by approximately $1$dB.  Furthermore, QuDeC outperformes the state-of-the-art method, UMRL+cycle-spinning~\cite{Authors19}, by 0.3dB on average. Figure~\ref{Fig:exp4} shows the qualitative performance of QuDeC against other methods on synthetic rainy images.  The DDN method is over de-raining on some images and on others it is slightly under de-raining as shown in the second column of Figure~\ref{Fig:exp4}. First four rows show under de-raining performance corresponding to methods RESCAN~\cite{Authors18d} and DIDMDN~\cite{Authors18} in the third and fourth columns of Figure~\ref{Fig:exp4} respectively where we can observe residue rain streaks in their outputs. The last three rows show over  de-raining of methods RESCAN~\cite{Authors18d} and DIDMDN~\cite{Authors18} in the third and the fourth columns of Figure~\ref{Fig:exp4}, respectively where we can observe the edges of the objects are blurred and objects like cables or wires and texture of birds feather have disappeared. Visually we can see in the fifth column of Figure~\ref{Fig:exp4} that our method produces images without any artifacts. From Figure~\ref{Fig:exp4} we can see that our method is able to 
	\begin{itemize}
		\item  recover the texture on the wooden wall in the first row,
		\item  produce clear objects with sharp edges in the fourth and the fifth images of the fifth coulmn,
		\item  remove rain streaks by maintaining the underlying textures on trees and on the  feathers in the third and the sixth images of the fifth coulmn.
\end{itemize}

	\begin{figure*}[htp!]
		\begin{center}
			\includegraphics[width=0.195\textwidth,height=0.12\textwidth]{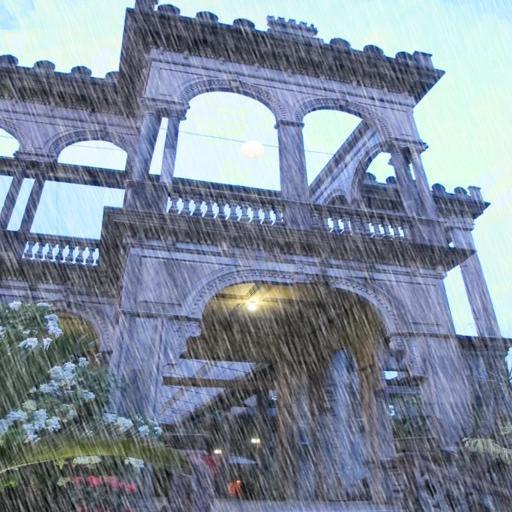}
			\includegraphics[width=0.195\textwidth,height=0.12\textwidth]{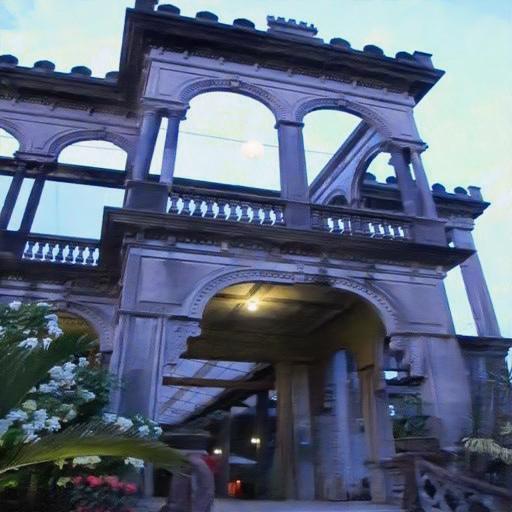}
			\includegraphics[width=0.195\textwidth,height=0.12\textwidth]{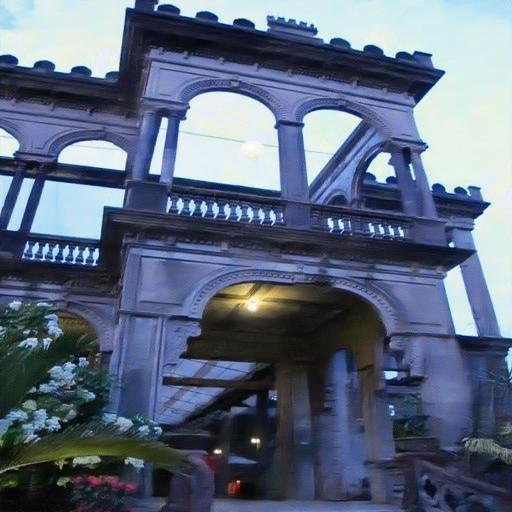}
			\includegraphics[width=0.195\textwidth,height=0.12\textwidth]{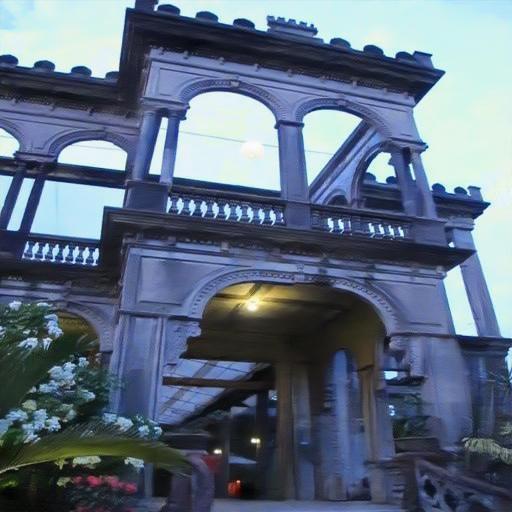}
			\includegraphics[width=0.195\textwidth,height=0.12\textwidth]{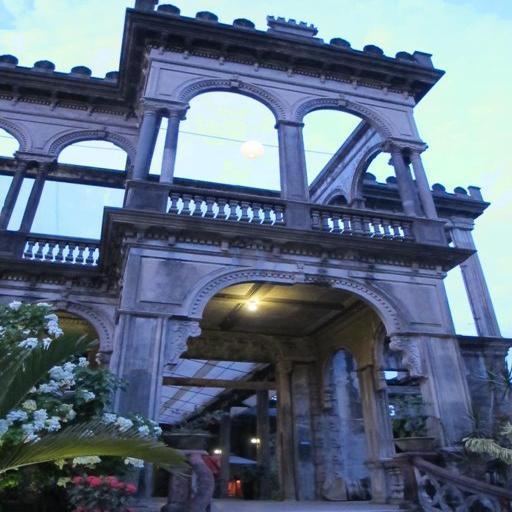}\\ \vskip3pt
			\includegraphics[width=0.195\textwidth,height=0.12\textwidth]{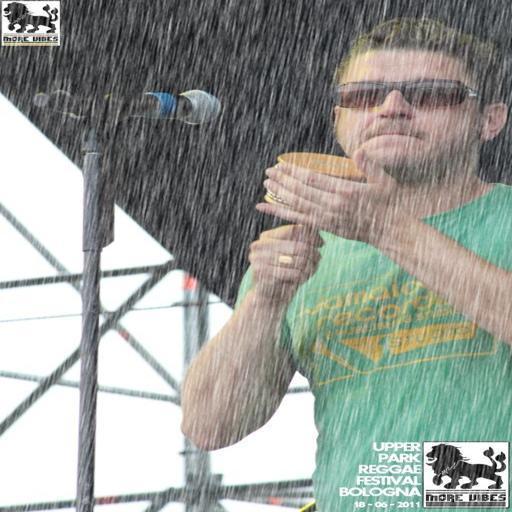}
			\includegraphics[width=0.195\textwidth,height=0.12\textwidth]{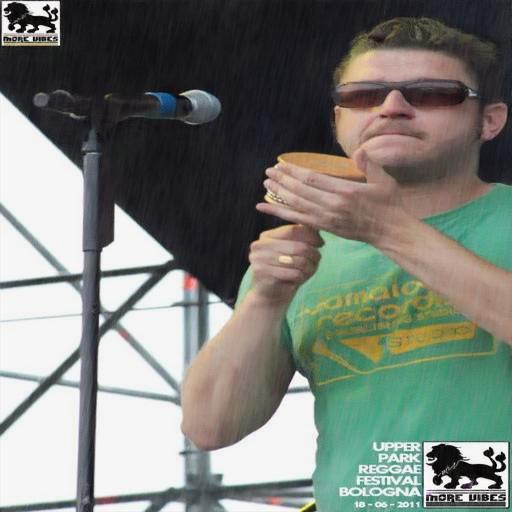}
			\includegraphics[width=0.195\textwidth,height=0.12\textwidth]{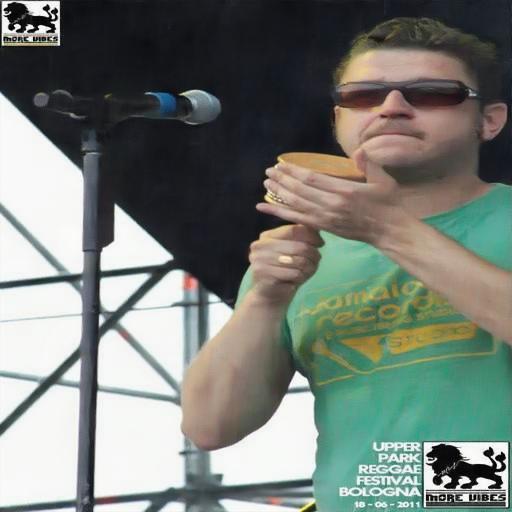}
			\includegraphics[width=0.195\textwidth,height=0.12\textwidth]{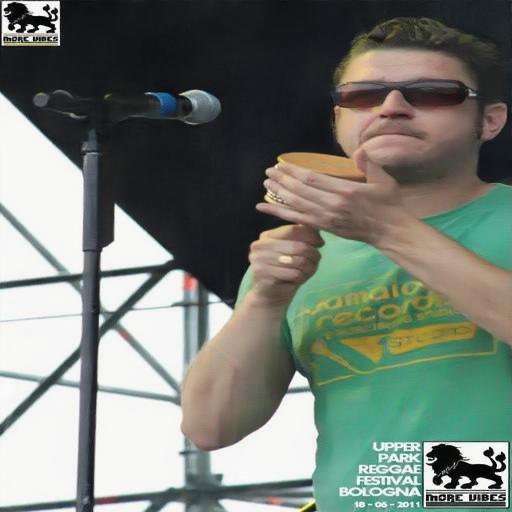}
			\includegraphics[width=0.195\textwidth,height=0.12\textwidth]{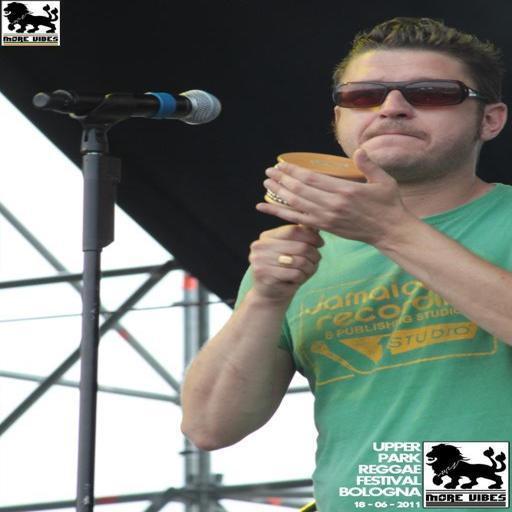}\\ \vskip3pt
			\includegraphics[width=0.195\textwidth,height=0.12\textwidth]{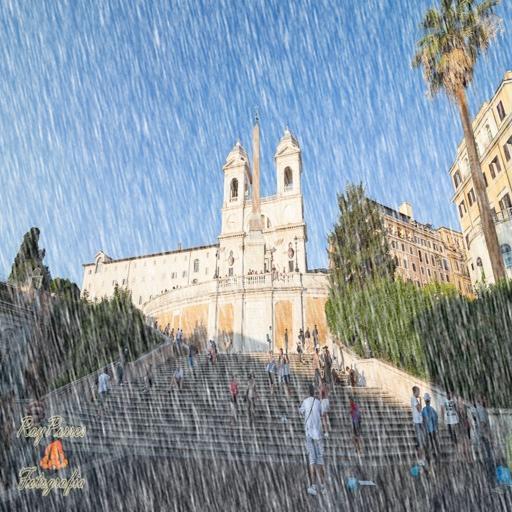}
			\includegraphics[width=0.195\textwidth,height=0.12\textwidth]{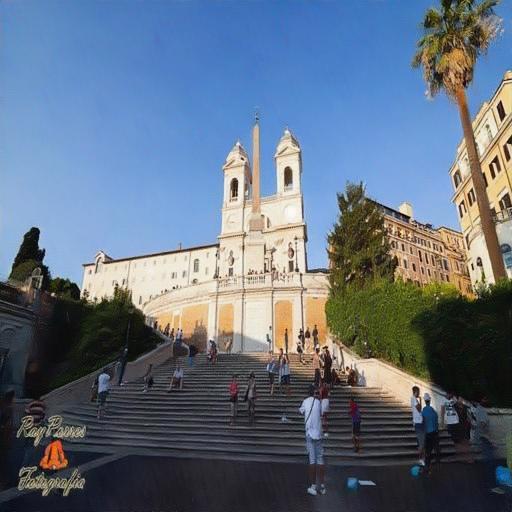}
			\includegraphics[width=0.195\textwidth,height=0.12\textwidth]{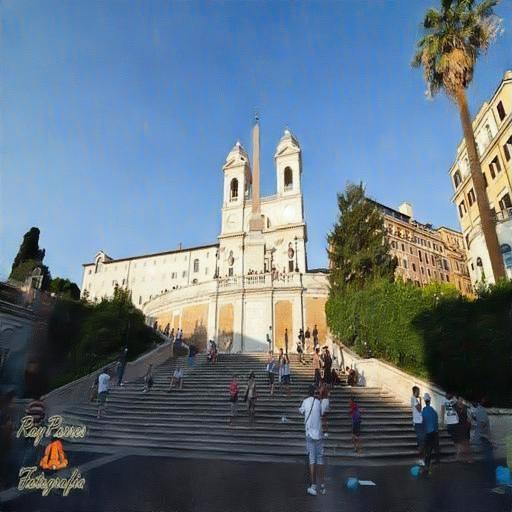}
			\includegraphics[width=0.195\textwidth,height=0.12\textwidth]{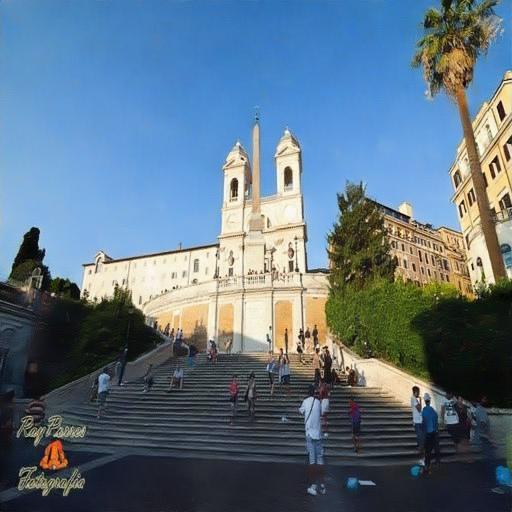}
			\includegraphics[width=0.195\textwidth,height=0.12\textwidth]{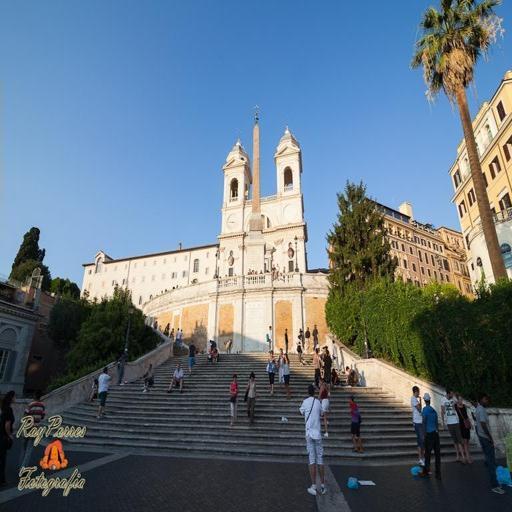}\\ \vskip3pt
			\includegraphics[width=0.195\textwidth,height=0.12\textwidth]{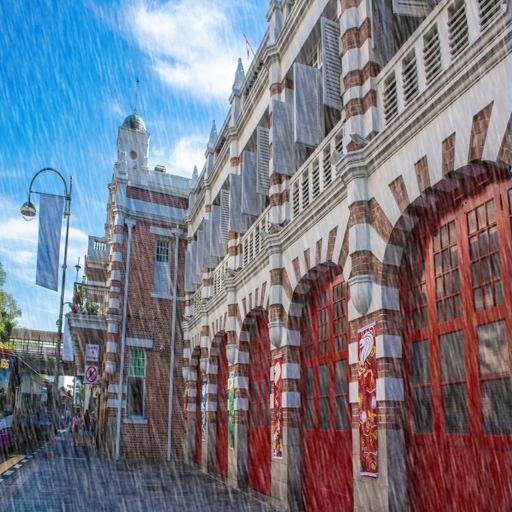}
			\includegraphics[width=0.195\textwidth,height=0.12\textwidth]{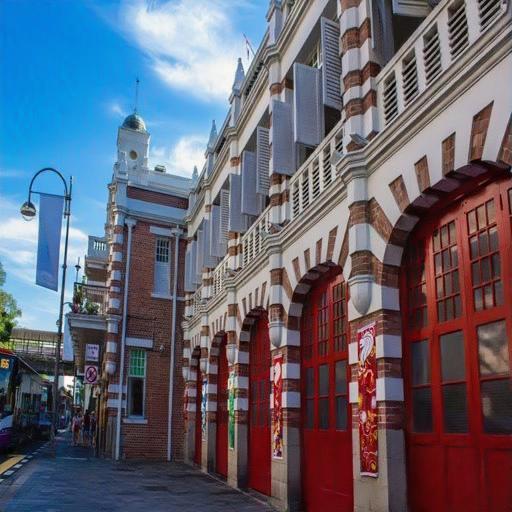}
			\includegraphics[width=0.195\textwidth,height=0.12\textwidth]{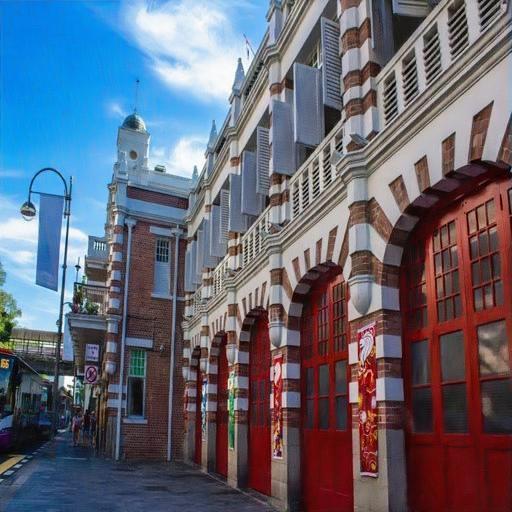}
			\includegraphics[width=0.195\textwidth,height=0.12\textwidth]{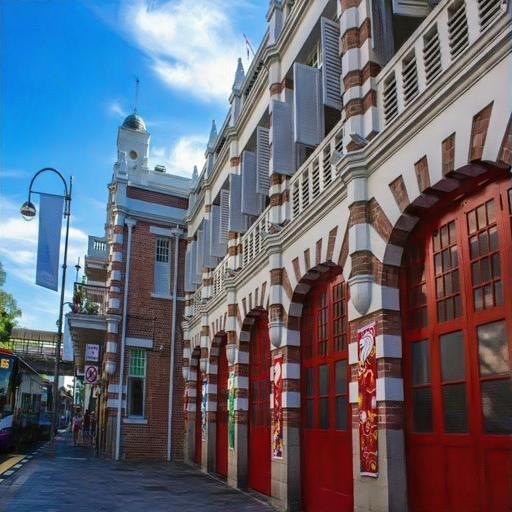}
			\includegraphics[width=0.195\textwidth,height=0.12\textwidth]{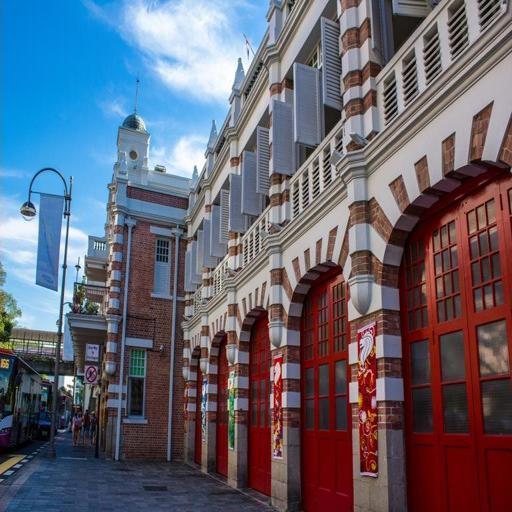}\\ 
			\textit{ Rainy  \hskip70pt Encoder- \hskip60pt QuDeC w/o  \hskip60pt QuDeC \hskip60pt Ground\\}
			\textit{Image \hskip80pt Decoder D1 \hskip80pt LCN \hskip60pt (Ours) \hskip60pt Truth \\ }
			\caption{Results of ablation study on synthetic images.}
			\label{Fig:exp2}
		\end{center}
	\end{figure*}
	
	\subsection{Results on Real-World Rainy Images}
	
	We conducted experiments on the real-world images provided by the authors of \cite{Authors17e,Authors17h,Authors18}. Results are shown in Figure~\ref{Fig:exp6}. Similar to the results obtained on synthetic images, we observe the same trend of either over de-raining or under deraining by the other methods. On the other hand, our method is able to remove rain streaks while preserving details of objects in the resultant output images. For example, the background texture on tree is sharp when compared to other methods. Also, rain streaks are removed properly without losing background information of walls and flowers in the second, third and fourth images of the fifth column in Figure~\ref{Fig:exp6}. All of these experiments clearly show that our method can handle different levels of rain (low, medium and high) with different shapes and scales.
	
	\subsection{Ablation study}
	We study the performance of each block’s contribution to the QuDeC network by conducting extensive experiments on the test datasets. We start with Encoder-Decoder D1 which are similar to UMRL ~\cite{Authors19} network. Encoder-Decoder D1 is trained as explained in UMRL~\cite{Authors19} method. Now we add Decoder D2, where we call the resultant network as QuDeC w/o LCN, and output of D2 is supervised with cross-entropy loss using the ground-truth maps, $s$. Finally, we add LCN to Decoder D2 to construct our proposed network QuDeC. Table~\ref{ablation} shows the contribution of each block on the performance of QuDeC network. Addition of Decoder D2, i.e formulating the joint task of computing distortion level at each location and estimating the residual rain streak component improves the overall performance by approximately 0.37dB. Furthermore, introducing LCN to Decoder D2 improves the performance of QuDeC by 0.2dB. Figure~\ref{Fig:exp2}, visually shows the improvement in performance after adding each block in constructing the QuDeC network. For example, we can clearly observe from  Figure~\ref{Fig:exp2}, QuDeC is able to reconstruct clear skies and dark backgrounds in the second and third images of the fourth column. Also QuDeC is able to reconstruct sharp objects when compared to the outputs of network with only Encoder-Decoder D1 in the first and last rows of Figure~\ref{Fig:exp2}.

	\begin{table}[htp!]
		\caption{PSNR and SSIM (PSNR$|$SSIM) results corresponding to the ablation study.}
		\label{ablation}
		\begin{tabular}{|c|c|c|c|c|}
			\hline
			Dataset & Testset  & \begin{tabular}[c]{@{}c@{}c@{}}UMRL~\cite{Authors19}\\ (Encoder-\\Decoder  D1)\end{tabular}       & \begin{tabular}[c]{@{}c@{}}QuDeC\\ w/o LCN\end{tabular}  & QuDeC      \\ \hline
			\multirow{2}{*}{DIDMDN~\cite{Authors18}} & \textit{Test-1}   & 29.42$|$0.91 & 30.17$|$0.92    & \textbf{30.43$|$0.93} \\ \cline{2-5} 
			& \textit{Test-2}   & 26.47$|$0.91 & 26.58$|$0.91    & \textbf{26.72$|$0.92} \\ \hline
			Rain800\cite{Authors17e} & \textit{Rain800}  & 24.19$|$0.85 & 24.42$|$0.86    & \textbf{24.61$|$0.86} \\ \hline
			JORDER~\cite{Authors17h} & \textit{Rain200H} & 26.17$|$0.91 & 26.56$|$0.92    & \textbf{26.74$|$0.93} \\ \hline
		\end{tabular}
	\end{table}
	
	\section{Conclusion}\label{sec:conclusion}
	We proposed a novel QuDeC to address the single image de-raining problem. In our approach, we formulate rain removal problem as a joint task of computing distortion level at each location and estimating the residual rain streak information. We judiciously combine the residual rain streak outputs at lower scales and distortion level information at each location using the corresponding confidence maps. Extensive experiments showed that QuDeC is robust enough to handle different levels of rain content for both synthetic and real-world rainy images.

	\appendices
	\section{Generation-of-Labels Network (GLN) Architecture}
	\begin{figure}[htp!]
		\centering
		\includegraphics[width=0.48\textwidth]{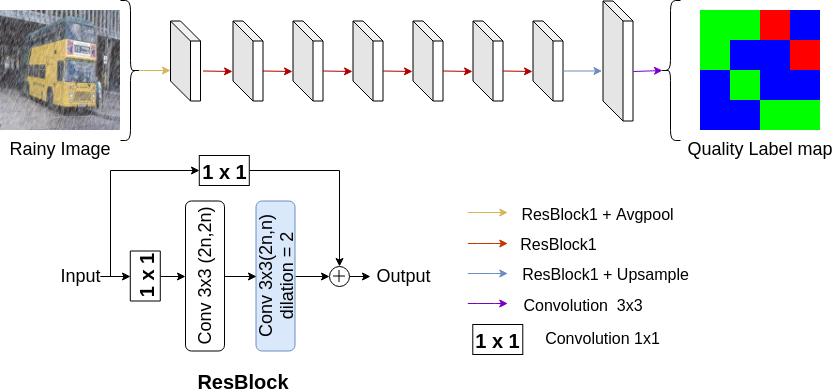}\\
		\caption{Generation-of-Labels Network (GLN).}
		\label{Fig:GLN}
	\end{figure}
	
	Generation-of-Labels Network (GLN) is used to generate the ground-truth location-quality-label-maps $s$. We use residual blocks (ResBlock) as our building module for the GLN network. GLN network consists of a sequence of eight ResBlocks as shown in Figure~\ref{Fig:GLN}. A ResBlock consists of a $1 \times 1$ convolution layer, a $3 \times 3$ convolution layer and a $3 \times 3$ convolution layer with dilation factor of 2 as shown in Figure~\ref{Fig:GLN}. Given $y$, the GLN network process each patch $p$ of size $128\times 128$ at a time and outputs the distortion quality-label $s(p)$ for the corresponding patch $p$. GLN is trained on the rainy image patches and the corresponding labels, $\{y,s^\dagger\}$. $s^\dagger$ is generated as explained in Section~\ref{sec:training}. GLN is trained for $40$ epochs using the cross entropy loss with the Adam optimizer and the learning rate is set equal to 0.0002.
	
	\section*{Acknowledgment}
	
	This research is based upon work supported by the Office of the Director of National Intelligence (ODNI), Intelligence Advanced Research Projects Activity (IARPA), via IARPA R$\&$D Contract No. 2019-19022600002. The views and conclusions contained herein are those of the authors and should not be interpreted as necessarily representing the official policies or endorsements, either expressed or implied, of the ODNI, IARPA, or the U.S. Government.

	\ifCLASSOPTIONcaptionsoff
	\newpage
	\fi

	
	
	%

	\bibliography{Derain_TIP19}
	\bibliographystyle{IEEEtran}

\end{document}